\documentclass[a4paper,11pt]{article}
\pdfoutput=1 
\usepackage{jinstpub} 
\usepackage{mathrsfs}
\usepackage{amsmath}
\usepackage{amssymb}
\usepackage{graphicx}
\usepackage{subfigure}
\usepackage{amssymb}
\usepackage{amsthm}
\usepackage{siunitx}
\usepackage{multirow}
\pdfoutput=1
\usepackage{pdfpages}
\DeclareGraphicsExtensions{.jpeg}

\title{\boldmath Numerical study of track distortion in the Large Prototype TPC with end-plate based on bulk Micromegas}

\author[a]{Deb Sankar Bhattacharya,}
\author[b,1]{Purba Bhattacharya,\note{Corresponding author.}}
\author[c,d]{Supratik Mukhopadhyay,}
\author[c,d]{Nayana Majumdar,}
\author[c,d]{Sandip Sarkar,}
\author[e,2]{Sudeb Bhattacharya,\note{Retired Senior Professor.}}
\author[f]{Paul Colas,}
\author[f]{David Atti{\'e},}
\author[f]{Serguei Ganjour,}
\author[g,2]{Aparajita Bhattacharya.}

\affiliation[a]{University of W{\"u}rzburg, Hubland Nord, Emil-Hilb-Weg 22, Physik West, W{\"u}rzburg 97074, Germany}
\affiliation[b]{Department of Physics, University of Cagliari and INFN, Strada prov.le per Sestu, km 1.00, 09042 Monserrato (CA), Italy}
\affiliation[c]{Applied Nuclear Physics Division, Saha Institute of Nuclear Physics, Kolkata - 700064, India}
\affiliation[d]{Homi Bhabha National Institute, BARC Training School Complex, Anushaktinagar, Mumbai, Maharashtra 400094, India}
\affiliation[e]{Applied Nuclear Physics Division, Saha Institute of Nuclear Physics, Kolkata - 700064, India}
\affiliation[f]{IRFU, CEA, Universit Paris-Saclay, F-91191 Gif sur Yvette, France}
\affiliation[g]{Department of Physics, Jadavpur University, Jadavpur, Kolkata - 700032, India}

\emailAdd{purba.bhattacharya85@gmail.com}

\abstract{The present $\mathrm{R}\&\mathrm{D}$ activities for the 
International Large Detector Time Projection Chamber (ILD-TPC) concern 
the adoption of 
the micro pattern devices for the gaseous amplification stage.
Seven Micromegas modules which are commissioned on the end-plate of 
a Large Prototype TPC (LPTPC) at DESY, were tested with a 5 GeV 
electron beam, under a 1 T magnetic field.
During experiments, reduced signal sensitivity as well as distortion
in the reconstructed track was observed at the boundaries of these 
modules.
Electrostatic field inhomogeneity near the module boundaries was 
considered to be the possible major reason behind these observations.
In the present work, this hypothesis has been explored using the 
Garfield simulation framework.
It has been possible to contain the computational complexity of the problem 
with suitable simplifications.
Qualitative and quantitative agreements with experimental results have 
been achieved.
Possibility of mitigating the problems has been proposed using the same 
simulation framework.}


\begin{document}
\maketitle
\flushbottom


\sisetup{range-phrase=--}

\section{Introduction}

\begin{figure}[hbt]
\centering
\includegraphics[scale=0.35]{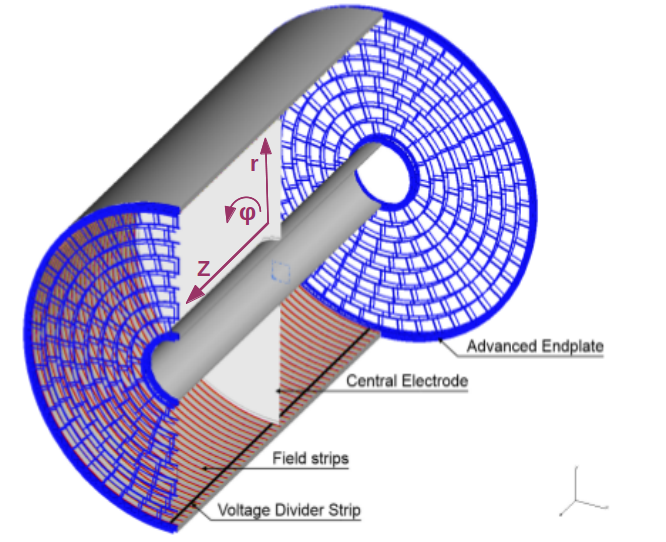}
\caption{Schematic of the TPC for the ILD \cite{ILC-TDR-Detector}.}
\label{TPC}
\end{figure}

At the International Linear Collider (ILC) \cite{ILC-TDR-Physics}, the 
electron-positron beam will collide at $250~\mathrm{GeV}$ in 
the center-of-mass frame.
A possible option of upgrading it to higher energies is also being considered.
The golden channel at $250~\mathrm{GeV}$ gives the unique opportunity of 
detecting Higgs events even without looking at its decay\cite{Fuji}.
Higgs mass can be determined with excellent precision from the measurement of 
Z boson, decaying in leptons.
The physics studies aimed at the ILC have pushed the requirements for
its detectors to an unprecedented level.
These requirements include excellent momentum resolution and good particle 
identification \cite{ILC-TDR-Detector}.
The International Large Detector (ILD) \cite{ILC-TDR-Detector} is one of the 
two detector concepts for the ILC.
A Time Projection Chamber (TPC) \cite{TPC, TPC-Hilke} (Fig.~\ref{TPC}) has 
been foreseen as the central tracker of the ILD and is likely to be installed 
just around the vertex detectors to accomplish continuous $3\mathrm{D}$ 
tracking with high efficiency.

The size of the planned ILD-TPC is $329~\mathrm{mm}$ for inner radius,
$1808~\mathrm{mm}$ for outer radius, and $2350\times2~\mathrm{mm}$
(divides two parts) in {\it{z}} direction \cite{ILC-TDR-Detector}.
The TPC is expected to be placed inside a magnetic field of $3.5~\mathrm{T}$.
Under these conditions, the ILD-TPC is projected to have the spatial 
resolution of less than \SI{100}{\micro\meter} and 2-hit resolution of 
less than $2~\mathrm{mm}$ \cite{ILC-TDR-Detector}.
The Linear Collider TPC (LCTPC) collaboration \cite{LCTPC} was formed to 
pursue the the design, development and test of a Large Prototype TPC (LPTPC) 
\cite{LPTPC, LPTPC-Peter-Scade}.

The Micro-Pattern Gaseous Detectors (MPGDs) \cite{MPGD}, due to their wide 
variety of geometries and flexible operating parameters, become a common 
choice for tracking and triggering detectors.
The LCTPC collaboration has, therefore, investigated the use of MPGDs for 
the TPC readout\cite{MPGD-TPC}.
One of the main advantage of the MPGD-based redaout over the conventional 
wire-based amplification system is the fact that in the amplification
 region, $\vec{E}\times\vec{B}$ effects are small and do not affect the 
resolution.

\begin{figure}[hbt]
\centering
\subfigure[]
{\label{Endplate}\includegraphics[scale=0.74]{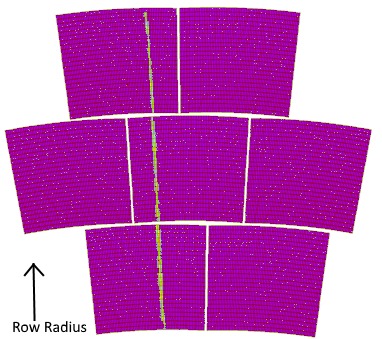}}
\subfigure[]
{\label{ModuleDetail}\includegraphics[scale=0.74]{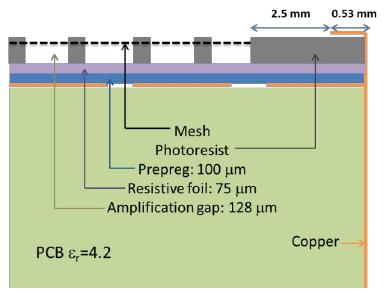}}
\caption{(a) A reconstructed track on the TPC end-plate made of seven Micromegas modules; (b) Schematic diagram of a region close to the edge of one of the module, side view \cite{PRF1}.}
\label{module}
\end{figure}

One of the most widely used MPGDs is Micro-MEsh Gaseous Structure (Micromegas) Detectors, invented by Y. Giomataris et al. \cite{Micromegas}.
Up to seven resistive Micromegas (MM) \cite{Resistive} modules have been studied at the
LPTPC end-plate since 2009 through 2015 and they have shown promising
performance as required for the ILD-TPC \cite{LCTPC1, LCTPC2}.
The keystone-shaped modules have identical size of 
$22\times17~\mathrm{cm^2}$ so as to fit in the end-plate.
The amplification gap and the micro-mesh wire pitch have been chosen to be
\SI{128}{\micro\meter} and \SI{63}{\micro\meter}, respectively.
Different types of resistive layers with surface resistance around
\SIrange[range-units=single]{3}{5}{~\mathrm{M\Omega/square}} have been tested at the LPTPC with an
electron beam of energy ranging from \SIrange[range-units=single]{1}{6}{~\mathrm{GeV}}, under a magnetic field of $1~\mathrm{T}$.
The anode readout of a Micromegas module is segmented in 1726 pads of size
$3\times7~\mathrm{mm^2}$ and arranged in 24 rows.
The pads are readout with the AFTER-based electronics
designed for the T2K experiment \cite{T2K}.
A thin copper frame that is connected to the resistive layer of the layout, is 
kept at ground potential.
The reconstruction of a typical track on the TPC end-plate, consisting of
seven modules, is shown in Fig.~\ref{Endplate}.
The close-up schematic diagram of the region near the module edge is shown in
Fig.~\ref{ModuleDetail} \cite{PRF1}.

\begin{figure}[hbt]
\centering
\subfigure[]
{\label{Expt1}\includegraphics[scale=0.37]{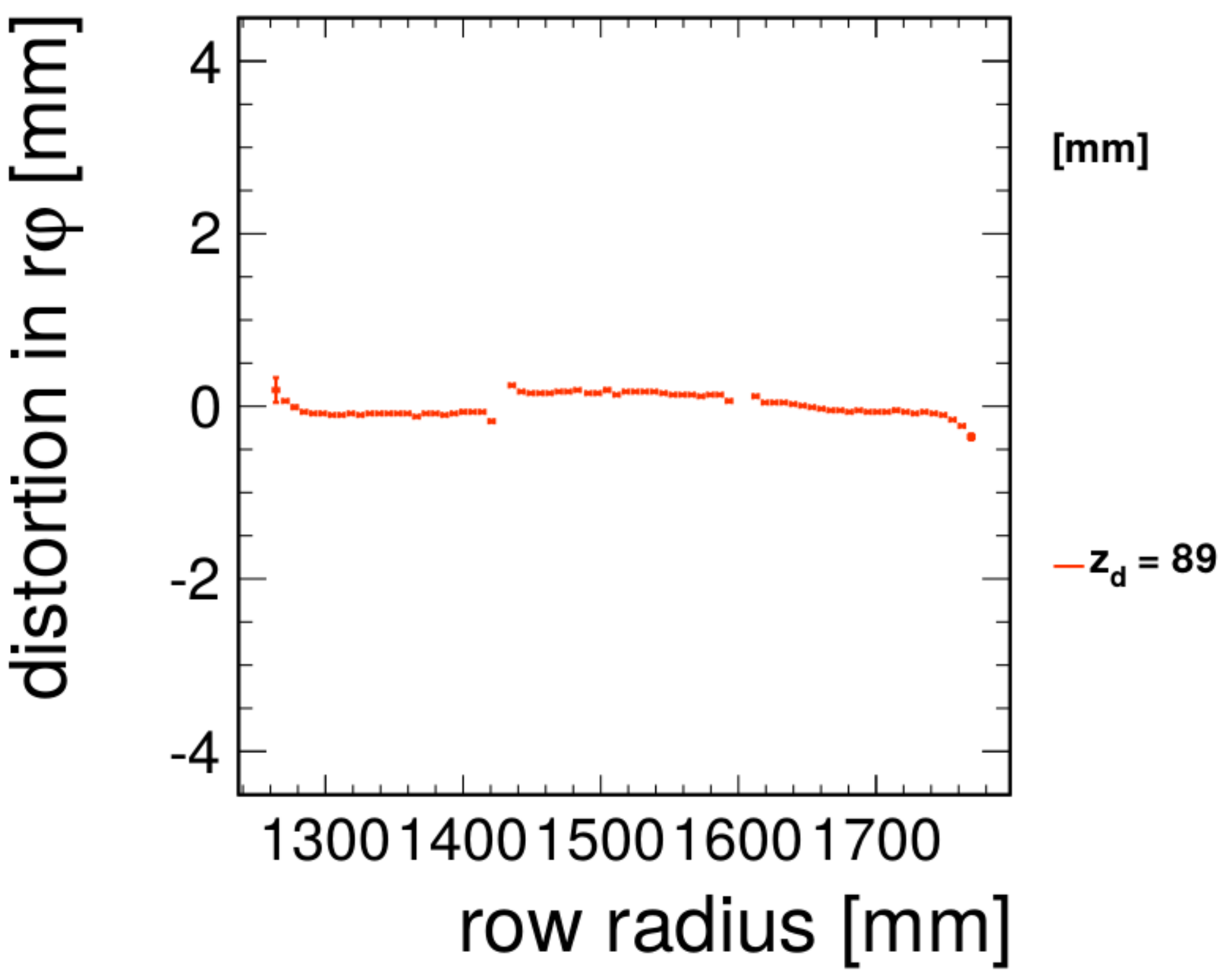}}
\subfigure[]
{\label{Expt2}\includegraphics[scale=0.37]{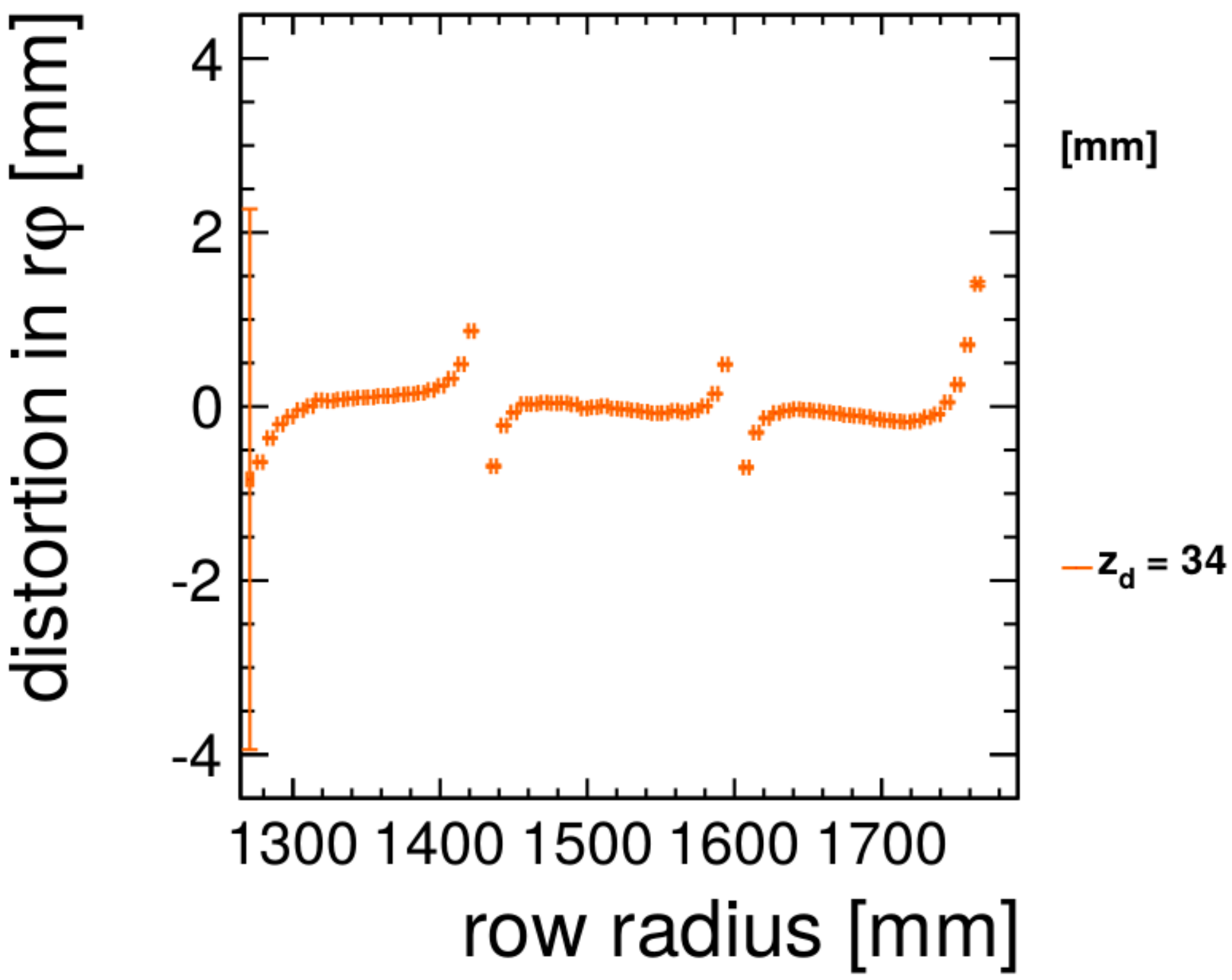}}
\caption{The experimentally observed distortion at (a) ${\it{B}}~\mathrm{= 0~T}$ (after the alignment correction), (b) ${\it{B}}~\mathrm{= 1~T}$ (no alignment correction) \cite{EPS-Deb}.}
\label{ExptResidue}
\end{figure}

During the analysis process, a reduced signal sensitivity was observed near
the edge of the module.
It was also found that the spatial resolutions of all the pads were consistent,
except for those near the module edge.
Figure~\ref{ExptResidue} shows the distortion (residual) plot versus the row
radius without or with magnetic field \cite{EPS-Deb}.
Residual of the pad hits on the extreme rows of the MM modules have
larger magnitude with respect to the other rows.
The transfer between two detector modules is seen and, in-between them, the
distortion is larger and leads to `S' like shape as shown in 
Fig.~\ref{Expt1} and Fig.~\ref{Expt2}.
It should be noted that the difference between the distortions is expected
since in the presence of magnetic field, the transverse diffusion is known to
be reduced and the distortion is likely to be influenced by
$\vec{{E}}~\times~\vec{{B}}$ effect.
The most likely reason leading to this distortion is the electric field
inhomogeneity introduced by the gap between a module and its neighbor.
Similar conclusion was obtained by studying the electric field configuration
near the module boundaries with GEM amplification stages using a
simplified version of the actual experimental setup \cite{Zenker, Zenker2}.
In what follows, besides investigating the electric field at the module
boundaries for an end-plate having Micromegas modules, we have numerically
estimated the residuals for a more realistic geometry and
compared our estimates with the experimental data.

\section{Numerical approach}

\subsection{Simulation Tools}

Garfield \cite{Garfield1, Garfield2} simulation framework has been used
in the following work.
The 3D electrostatic field simulation has been carried out using neBEM
\cite{neBEM1, neBEM2, neBEM3} toolkit. Besides neBEM, 
HEED \cite{HEED1, HEED2} has been used for primary
ionization calculation and
Magboltz \cite{Magboltz1, Magboltz2} for computing drift, diffusion,
Townsend and attachment coefficients.

\subsection{Optimization of numerical model}

\begin{figure}[hbt]
\centering
\includegraphics[scale=0.8]{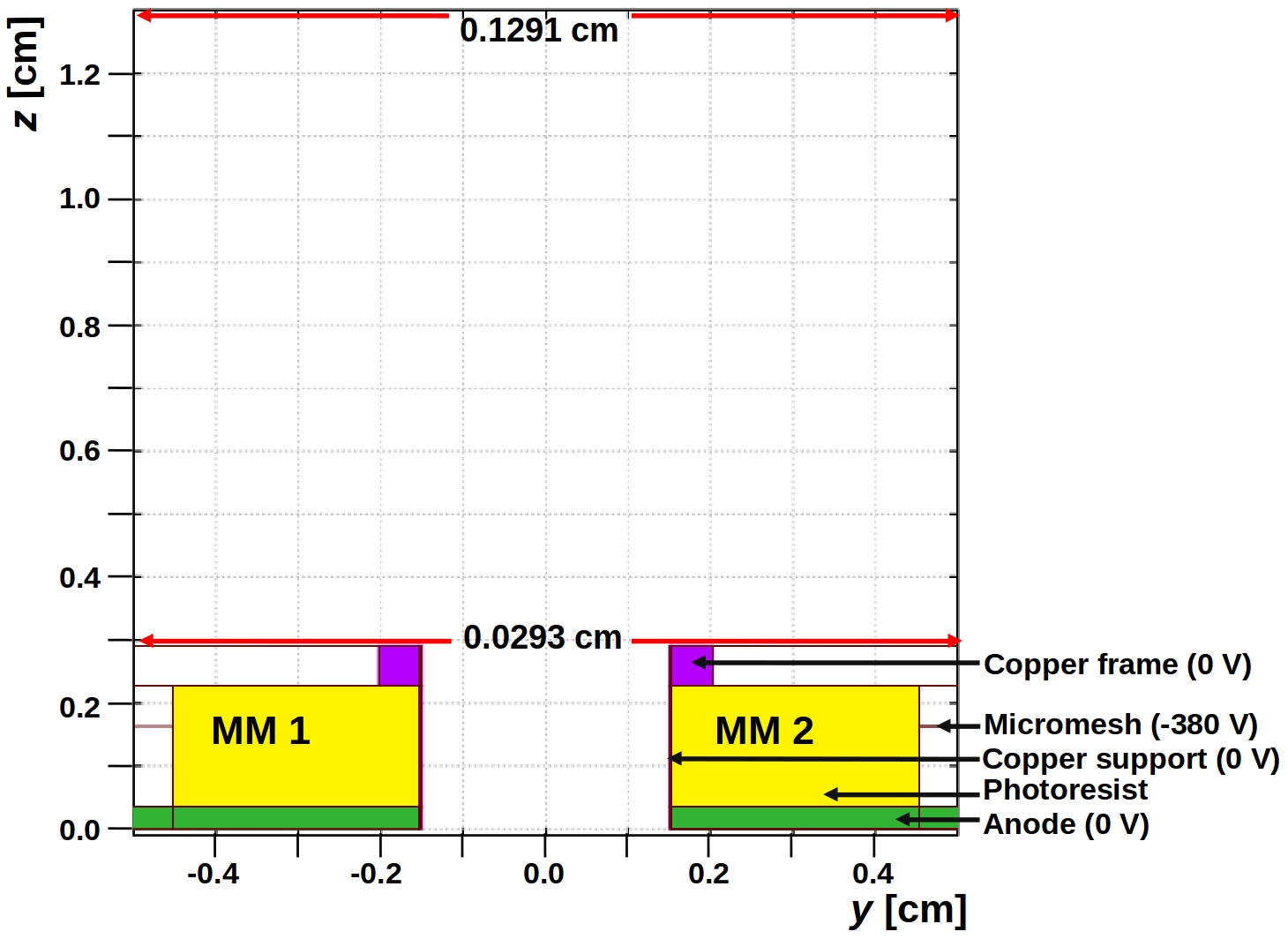}
\caption{Detailed side view of a numerical module.}
\label{Area-Case1}
\end{figure}

Instead of considering all seven $22\times17~\mathrm{cm^2}$
Micromegas modules for simulation, an attempt has been made to optimize
computational expenses
in terms of both number and size of the modules in the numerical model. 
The aim has been to minimize computation retaining the essential physics 
issues in the solution. 
The cross-sectional view of the simulated geometry is shown in
Fig.~\ref{Area-Case1}  where different parts of the geometry are explained.

\begin{figure}[hbt]
\centering
\subfigure[]
{\label{Area-2}\includegraphics[height=0.25\textheight]{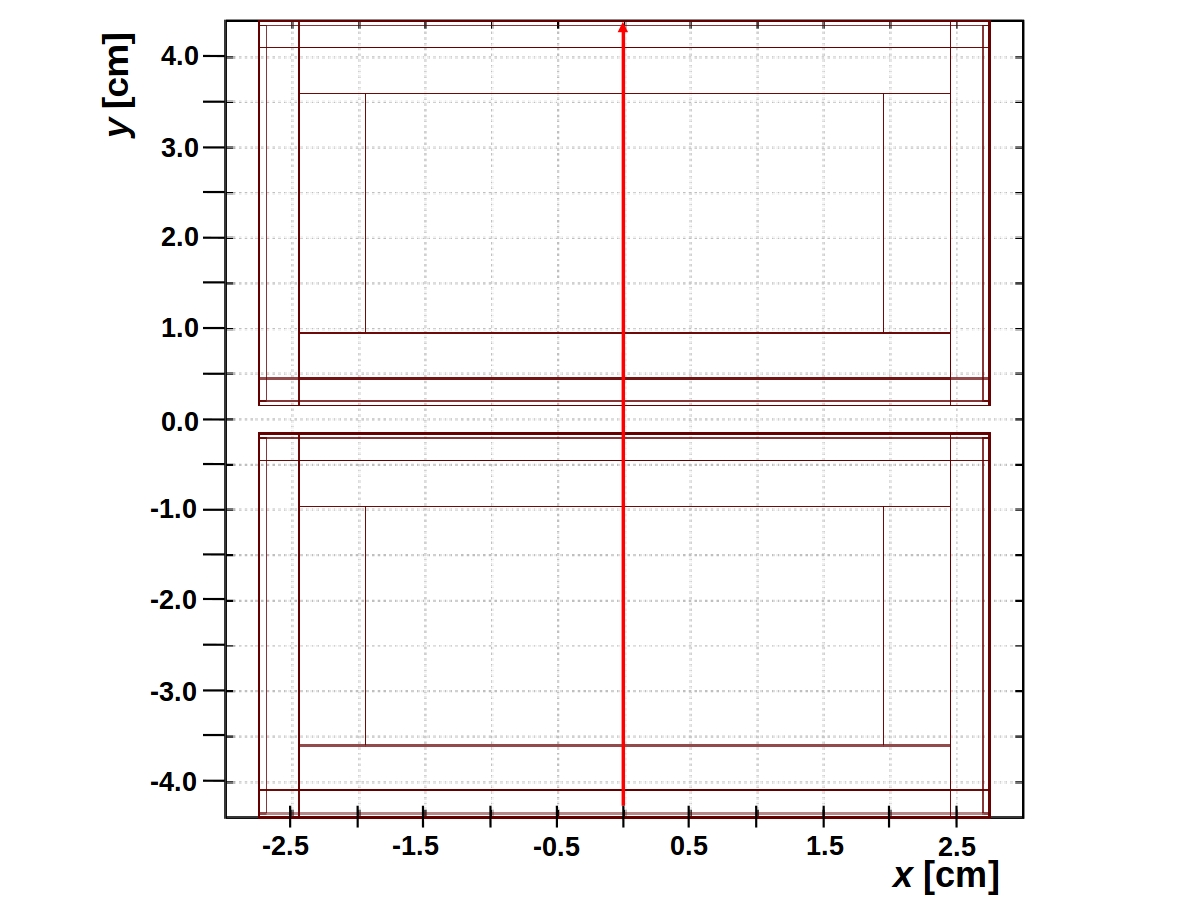}}
\subfigure[]
{\label{Area-3Modules}\includegraphics[height=0.25\textheight]{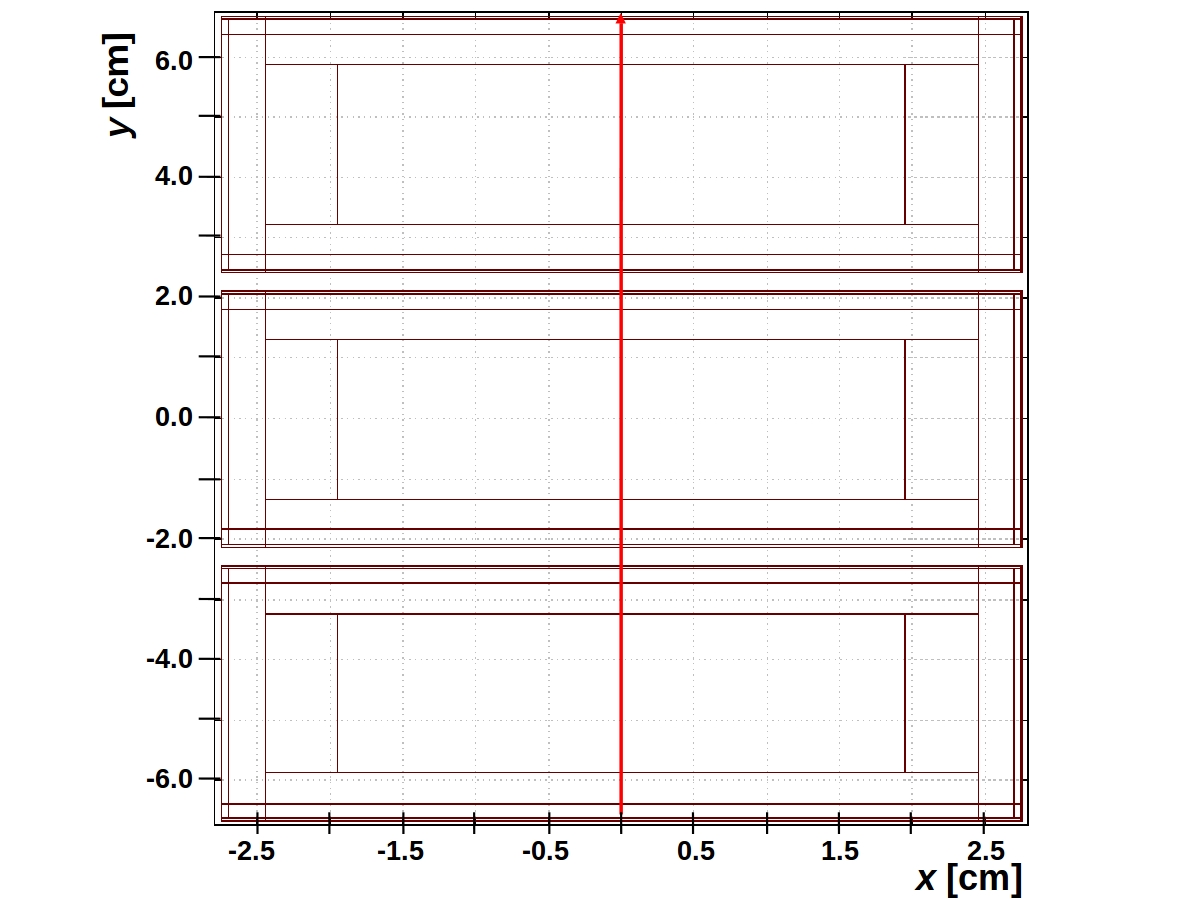}}
\caption{(a) Top view of the geometry (a) with two modules, (b) with three modules.}
\label{Area}
\end{figure}

For this part of the work, two Micromegas modules of size
$22\times17~\mathrm{cm^2}$ have been placed side by side,
as shown in Fig.~\ref{Area-2} and we have focused our attention on the
variation of potential and electric field close to the vicinity of the gap
between the two modules.
Similarly, since the track distortion studies carried out in this work
involves the effect of the inter-modular space on track reconstruction,
only three modules have been considered as shown in Fig.~\ref{Area-3Modules}.
 This configuration allows two inter-modular gaps to appear in the model
which is also typical in the experimental scenario (Fig.~\ref{ExptResidue}).
It may also be mentioned here that while the gap between any two modules is
$\mathrm{3~mm}$, the grounded copper extends to $\mathrm{0.53~mm}$ and the
photoresist is $\mathrm{2.5~mm}$ wide, as shown in Figs.~\ref{ModuleDetail}
and \ref{Area-2}.
Please note that the mesh begins at a distance of $\mathrm{4.53~mm}$ from the mid-point
between two modules.
Finally, it should be noted that for the 2-module geometry,
${\it{y}}~\mathrm{=0.0~cm}$ is the mid-point between the two modules, while for
the 3-module geometry, ${\it{y}}~\mathrm{=-2.275~cm}$ and $\mathrm{+2.275~cm}$ are
the two mid-points among the three modules along {\it{y}}.

\begin{figure}[hbt]
\centering
\subfigure[]
{\label{Solver-EY-Run2}\includegraphics[height=0.225\textheight]{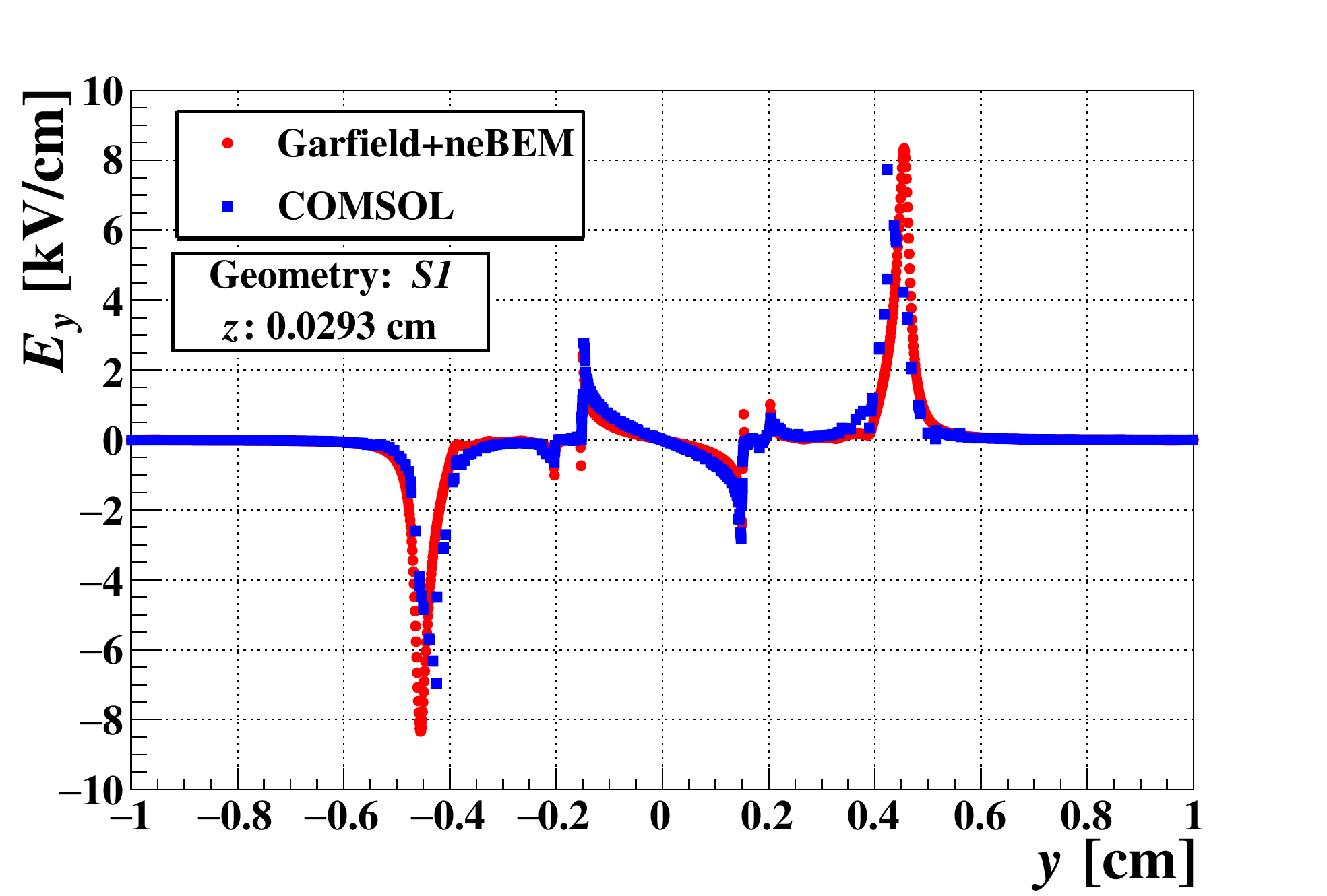}}
\subfigure[]
{\label{Solver-EZ-Run2}\includegraphics[height=0.225\textheight]{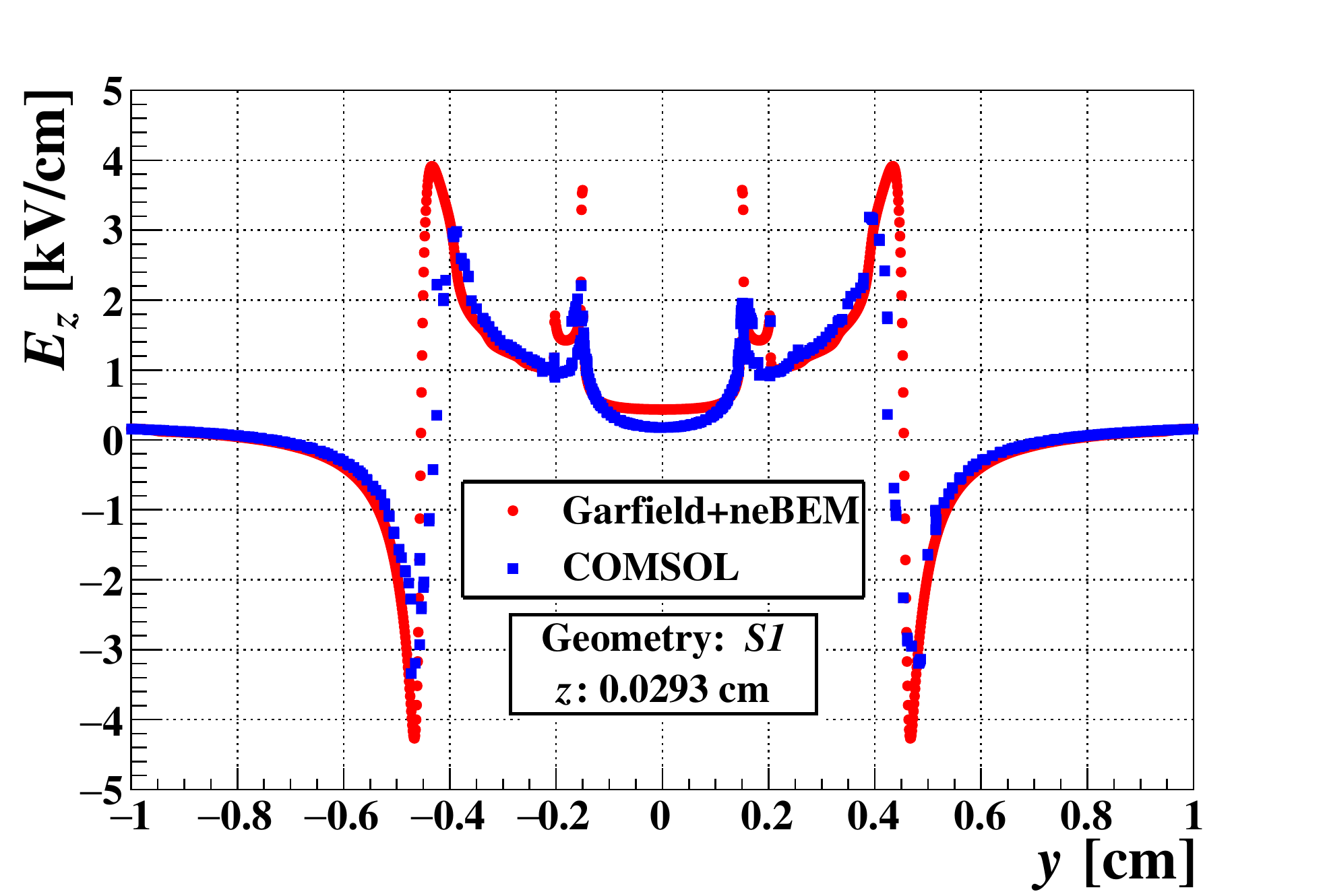}}
\caption{(a) $\it{E}_{\it{y}}$ and (b) $\it{E}_{\it{z}}$ along $\it{z} = \mathrm{0.0293}~\mathrm{cm}$ using COMSOL and neBEM. Micromegas modules have size of $22\times17~\mathrm{cm^2}$.}
\label{Solver-Run2}
\end{figure}

\begin{figure}[hbt]
\centering
\subfigure[]
{\label{Solver-EY-Run3}\includegraphics[height=0.225\textheight]{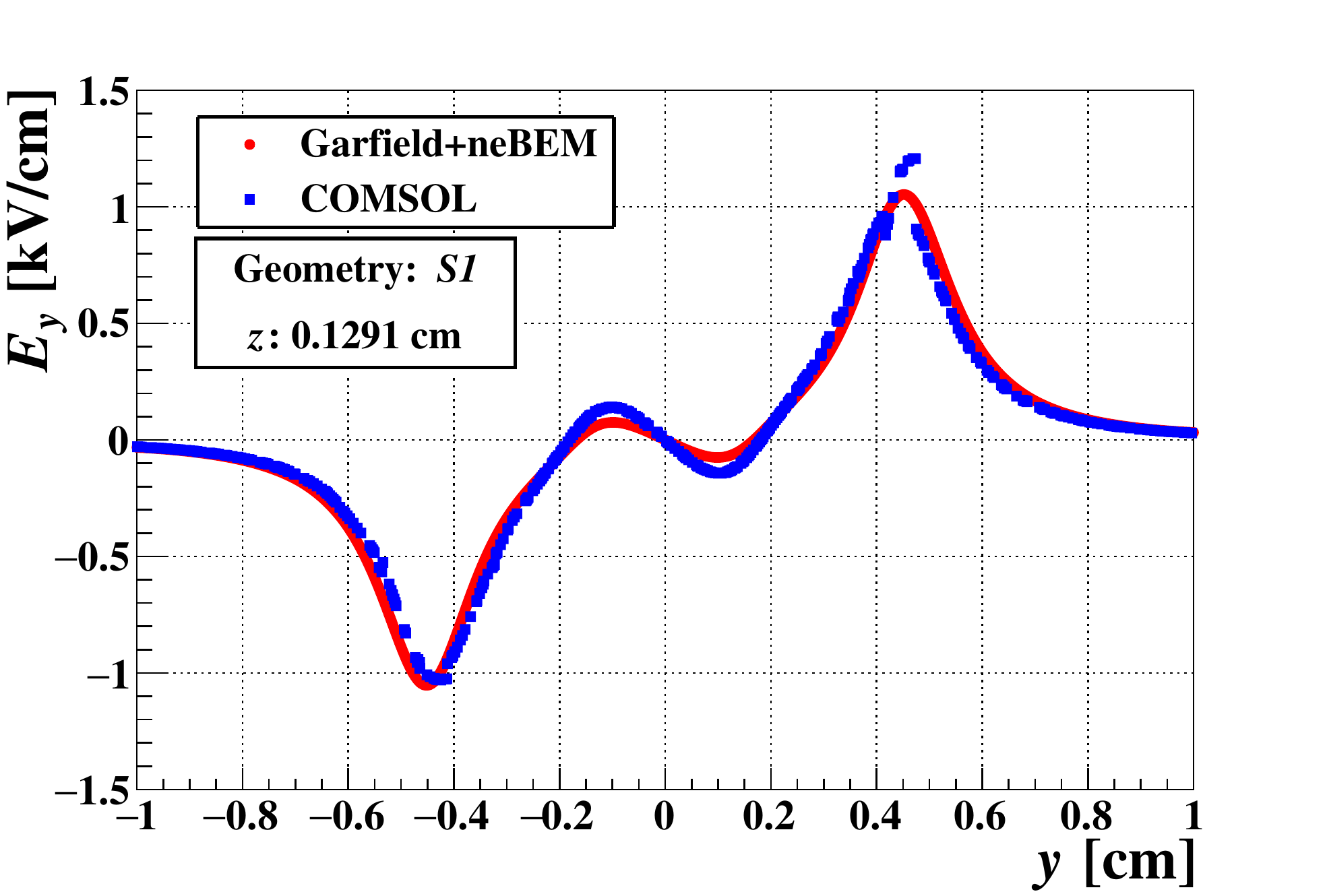}}
\subfigure[]
{\label{Solver-EZ-Run3}\includegraphics[height=0.225\textheight]{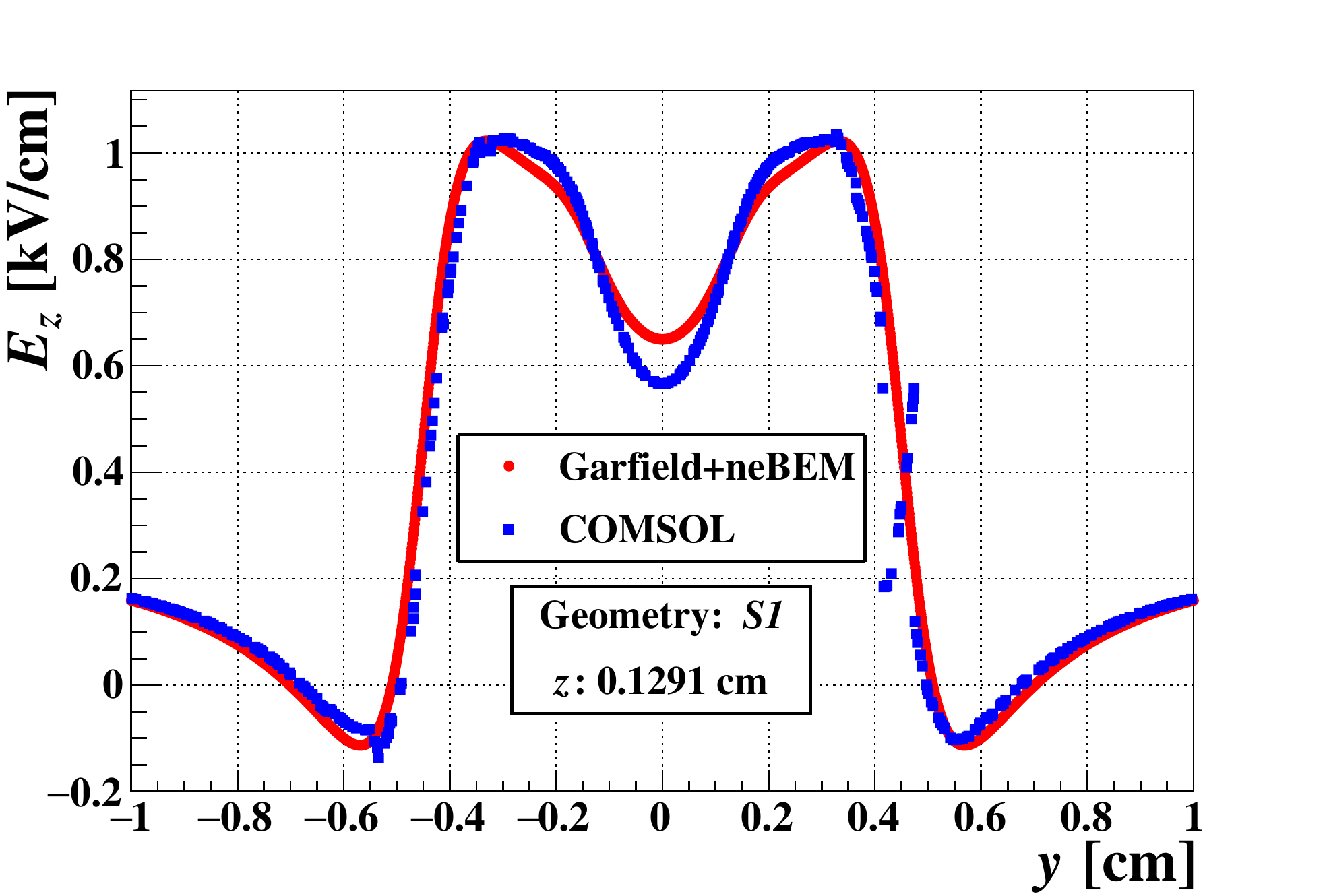}}
\caption{(a) $\it{E}_{\it{y}}$ and (b) $\it{E}_{\it{z}}$ along $\it{z} = \mathrm{0.1291}~\mathrm{cm}$ using COMSOL and neBEM. Micromegas modules have size of $22\times17~\mathrm{cm^2}$.}
\label{Solver-Run3}
\end{figure}

Two different tools which rely on completely different 
mathematical foundations have been 
compared to ensure the rationality of the numerical estimates. 
The Finite Element Method (FEM) based commercial solver COMSOL Multiphysics
\cite{Comsol} and the neBEM + Garfield combinations have been used for this 
purpose.
It should be mentioned here that the geometry of the overall device poses an 
extremely difficult problem for the both solvers. 
The component lengths span over several orders of magnitude. For example, 
length of a module is $\mathrm{22~cm}$ whereas the copper frame width is 
\SI{30}{\micro\meter} (7000:1). 
The situation is even worse if the entire device is considered. 
Despite the difficulties, an effort was made to maintain accuracy of the 
results and in Figs.~\ref{Solver-EY-Run2}, \ref{Solver-EZ-Run2}, 
\ref{Solver-EY-Run3} and \ref{Solver-EZ-Run3}, comparison between the two 
electric field solvers have been presented.
Despite the difference in mathematical models and numerical implementation of
the solution procedure, the estimates agree with each other in all broad
aspects.
Away from the gap, the field values are found to be uniformly distributed but
near the gap the transverse electric field is sharply increasing from
$\mathrm{0~ kV/cm}$ (nominal value) to $\pm\mathrm{8~kV/cm}$, while the
axial field is found to rise from the nominal value of around
$\mathrm{0.23~kV/cm}$ to $\pm\mathrm{4~kV/cm}$.
Noticeable non-uniformity of the electric field along {\it{y}}  of the
geometry is found to be extended up to nearly $\mathrm{1~cm}$ from the centre
of the gap, i.e. $\mathrm{4~mm}$ within the region where the micro-mesh is
known to exist.
Beyond this region of non-uniformity, all the fields are found to attain and
maintain their nominal values.

There are differences in magnitudes between the electric fields estimated by
different approaches.
The maximum disagreement is approximately 20\% in the most severe case, as
shown in Fig.~\ref{Solver-EZ-Run2}, but all the characteristic features
are found to be repeated using both the approaches.
From this study, we conclude that the electric field estimated by the
Garfield + neBEM combination are accurate and use this framework for further
studies.

According to the present understanding, the distortion is driven by 
the non-uniformity of the electrostatic field at the module boundaries. 
Thus, the reduction in size has been pursued as long as the features of the 
electrostatic field at the edges remain unaltered from larger modules.
As far as electrostatic non-uniformity is concerned, the geometry with the
chosen smaller modules is expected to be representative of the
real end-plate used in the test beams.
A brief numerical experiment has been performed to decide 
the optimum size of  the module in the numerical model, as described below.
For this part of the work, we have considered four possible sizes of 
MM modules which as shown in Table~\ref{table1}. 

\begin{table}[hbt]
\centering
\begin{tabular}{|c|c|}
\hline
Geometry & Size \\
\hline
{\it{S1}} & $22\times17~\mathrm{cm^2}$ \\
\hline
{\it{S2}} & $11\times8.5~\mathrm{cm^2}$ \\
\hline
{\it{S3}} & $5.5\times4.25~\mathrm{cm^2}$ \\
\hline
{\it{S4}} & $2.75\times2.125~\mathrm{cm^2}$ \\
\hline
\end{tabular}
\caption{Size of different MM modules used in the simulation}
\label{table1}
\end{table}

\begin{figure}[hbt]
\centering
\subfigure[]
{\label{Size-EY-Run2}\includegraphics[height=0.225\textheight]{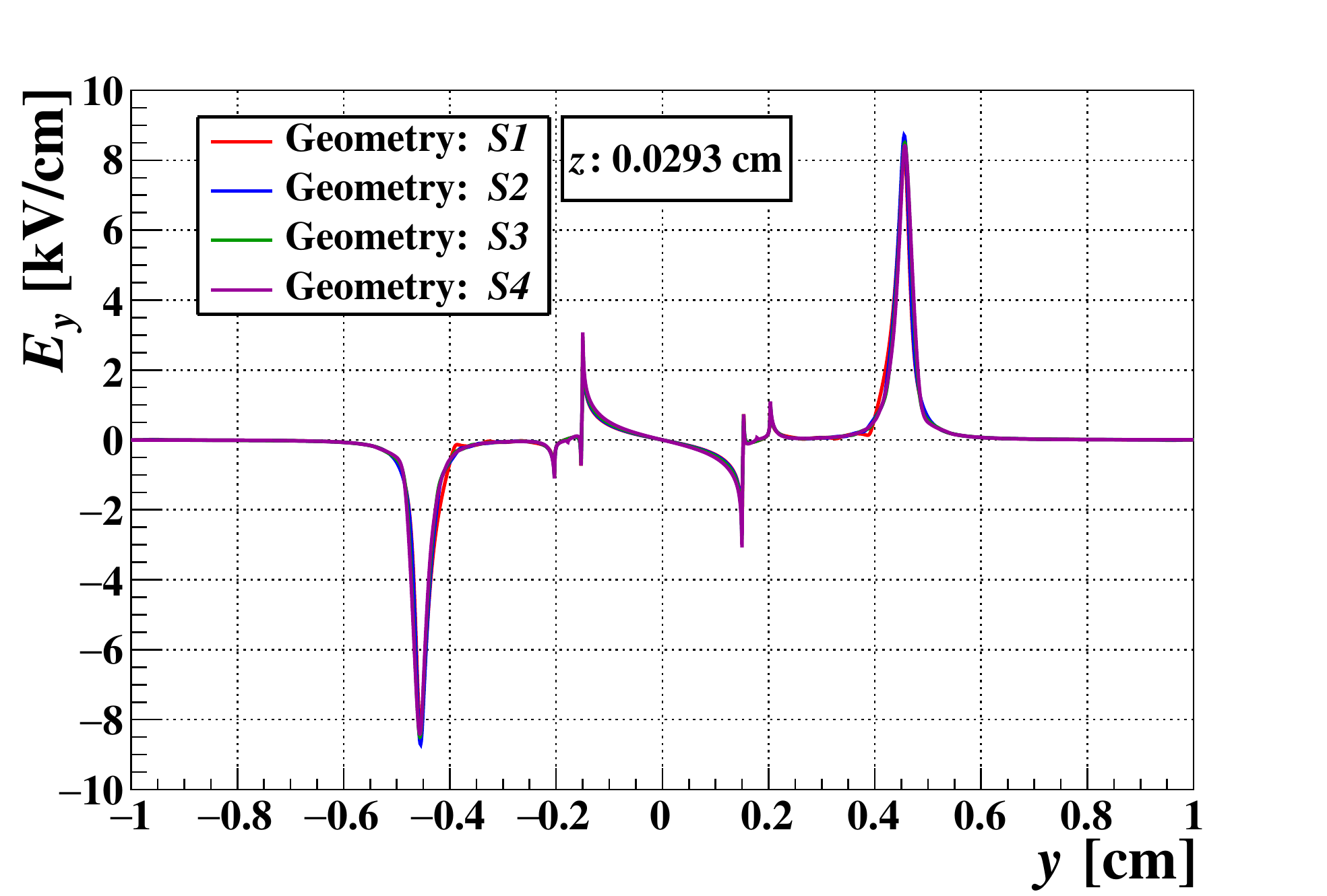}}
\subfigure[]
{\label{Size-EZ-Run2}\includegraphics[height=0.225\textheight]{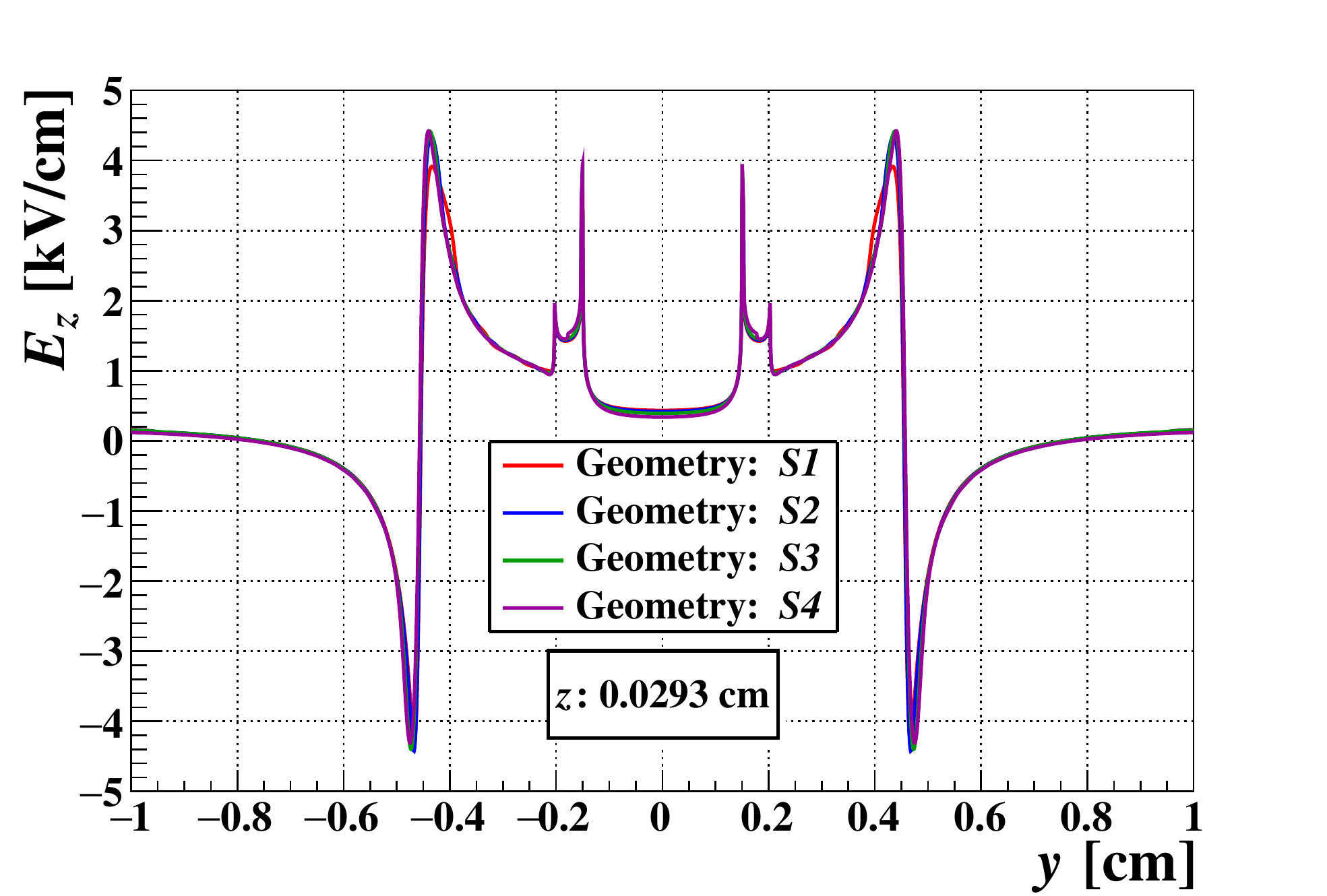}}
\caption{(a) $\it{E}_{\it{y}}$ and (b) $\it{E}_{\it{z}}$ along $\it{z} = \mathrm{0.0293}~\mathrm{cm}$ for geometries with several module sizes.}
\label{Size-Run2}
\end{figure}

\begin{figure}[hbt]
\centering
\subfigure[]
{\label{Size-EY-Run3}\includegraphics[height=0.225\textheight]{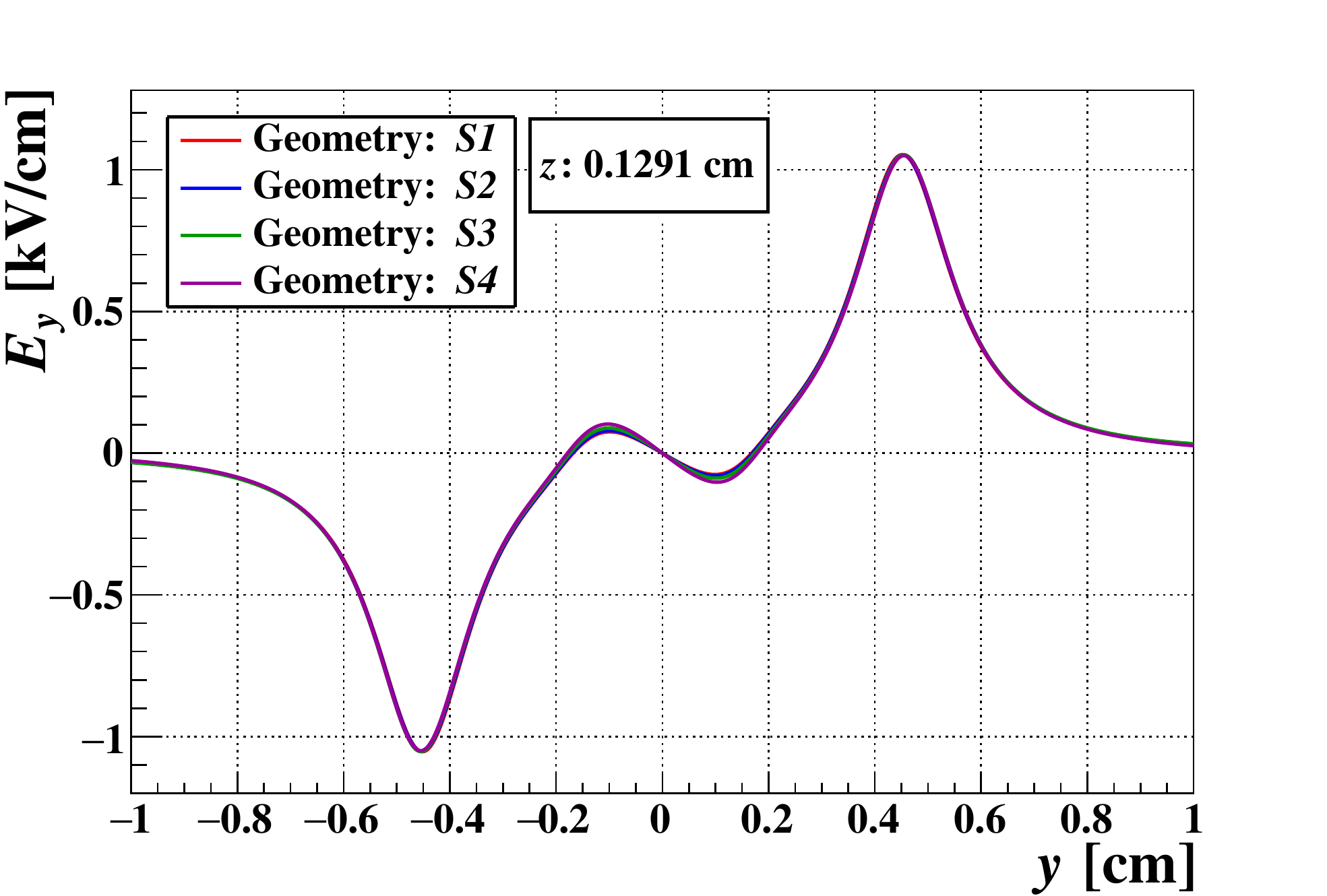}}
\subfigure[]
{\label{Size-EZ-Run3}\includegraphics[height=0.225\textheight]{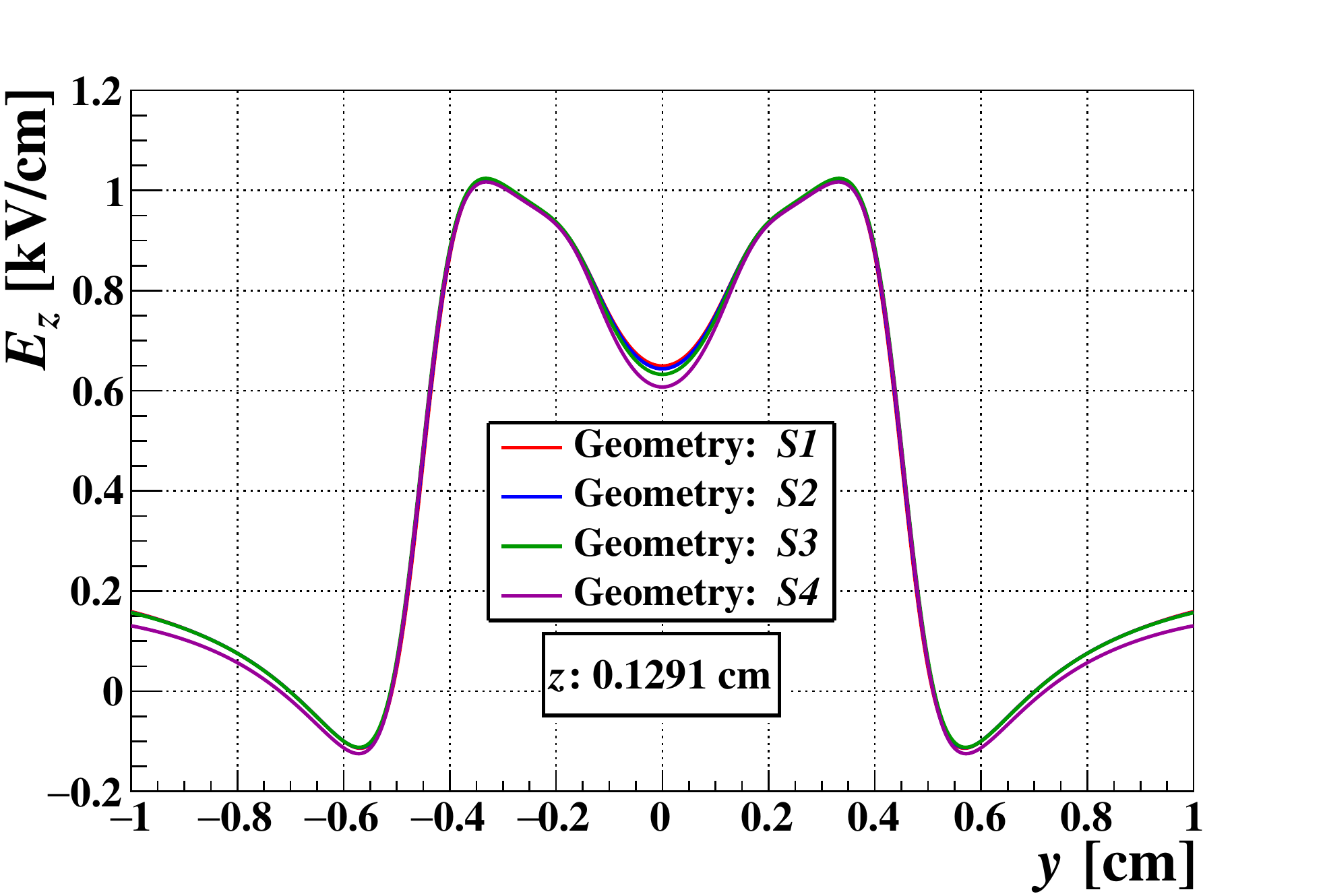}}
\caption{(a) $\it{E}_{\it{y}}$ and (b) $\it{E}_{\it{z}}$ along $\it{z} = \mathrm{0.1291}~\mathrm{cm}$ for geometries with several module sizes.}
\label{Size-Run3}
\end{figure}

We present the effect of module size on the field near the module boundary.
In Figs.~\ref{Size-EY-Run2} and \ref{Size-EY-Run3},
${\it{E_{y}}}$ estimates have been compared close to the module boundary (the zero
value coincides with the mid-point between the two modules) at
$\mathrm{0.0293~cm}$ above the anode plane (which is equivalent to a micron
above the top-most surface of a given module) and $\mathrm{0.1291~cm}$ above
the anode plane.
It can be seen that the estimated field values are identical for all the four
sizes of the numerical modules.
Similarly, ${\it{E_{z}}}$ values have been compared in Figs.~\ref{Size-EZ-Run2}
and \ref{Size-EZ-Run3}.
While Fig.~\ref{Size-Run2} shows excellent agreement among the different
sizes, Fig.~\ref{Size-Run3} indicates that ${\it{E_{z}}}$ for the smallest numerical
model is different from bigger three sizes.
From this study, we conclude that except the smallest numerical model, the
other three models produce estimates of electric fields that are consistent
with each other.
In order to reduce computational expenses, the smallest one among the allowed
three, the ${\mathrm{5.5}\times\mathrm{4.25~cm^2}}$ model, is chosen for
further studies.

Although the modelled geometry used in the present calculations is smaller and 
simpler than that of the real Micromegas on the LPTPC end-plate in terms of 
size and number, all critical parameters (e.g., the thickness of the copper 
frame and the photoresist wall, gap between the modules) which influence the 
electric field near the gap are given realistic values.
By comparing the experimental condition, as depicted in Fig.~\ref{module}, 
and the numerical model shown in Fig.~\ref{Area} and Fig.~\ref{Area-Case1},
 the following may be pointed out:
\begin{itemize}
\item{The actual experiment was performed using a resistive bulk Micromegas,
whereas, for this simulation, the standard bulk Micromegas has been
considered.}
\item{Instead of mesh, zero-thickness plane has been used to model the
micro-mesh plane.}
\item{A continuous grounded anode plane has been placed \SI{128}{\micro\meter} below the mesh plane. }
\item{To maintain a uniform drift field, the drift plane has been placed
$1~\mathrm{cm}$ above the modules.}
\item{As in the experiment, $3~\mathrm{mm}$ thick photoresist has been used to
support the modules.}
\item{True to the experiment, a copper layer has been connected all around the
module to the ground.}
\item{Between the modules there is a gap of $3~\mathrm{mm}$ which follows the
experimental situation.}
\end{itemize}

It should also be mentioned that, for the results presented here, the
micro-mesh plane has been biased with a potential of $-380~\mathrm{V}$
whereas $-610~\mathrm{V}$ has been applied to the drift plane, thus
creating a drift field of $230~\mathrm{V/cm}$. 
In all the cases, true to the experiment, a magnetic field of either 
$0~\mathrm{T}$ or $1~\mathrm{T}$ has been applied.
For the remaining calculations presented below, $\mathrm{T2K}$ gas
($\mathrm{Argon}$ $95\%$, $\mathrm{CF_4}$ $3\%$, $\mathrm{Isobutane}$ $2\%$)
has been considered.
The track along which the calculations have been performed are shown in
Fig.~\ref{Area}.
It should be mentioned here that while {$r$-$\phi$-$z$} coordinate 
system has been used in the experimental data interpretation, the numerical 
model utilizes the Cartesian {$x$-$y$-$z$} coordinate system. 
The {$r$-$\phi$} in experiment corresponds to {\it{x}} in numerical 
model and, similarly, $r$ in experiment corresponds to {\it{y}} 
in simulation. 

\section{Results}

\begin{figure}[hbt]
\centering
\subfigure[]
{\label{Drift-1}\includegraphics[height=0.3\textheight]{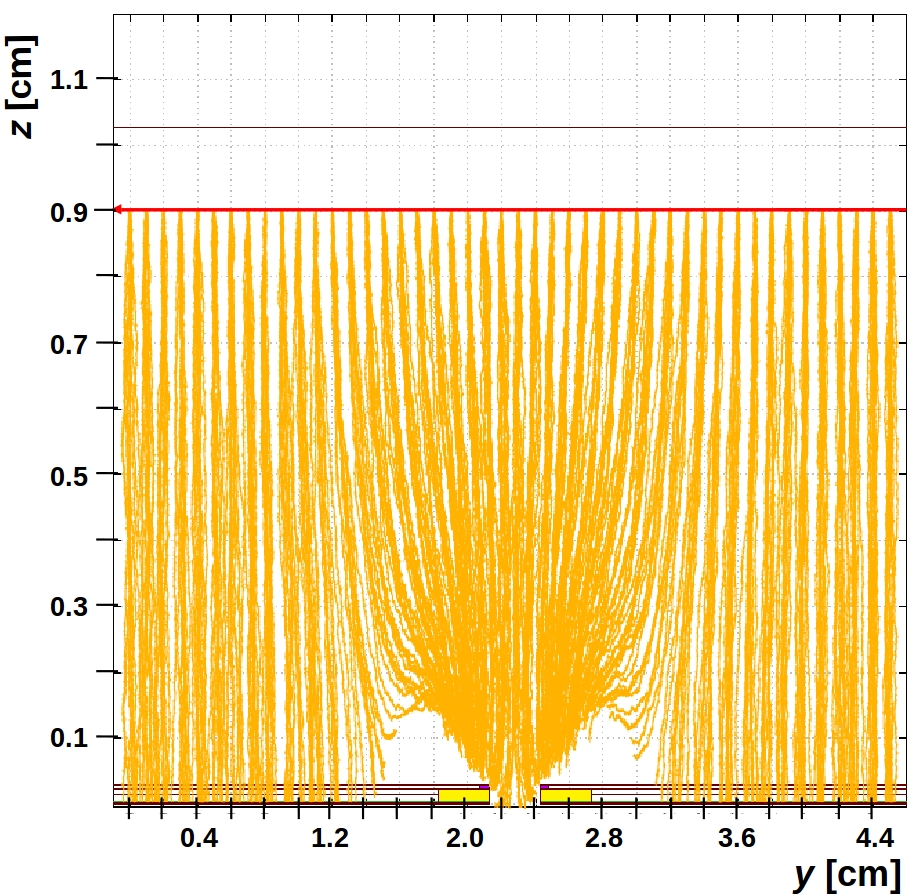}}
\subfigure[]
{\label{Drift-YView-Case1}\includegraphics[height=0.3\textheight]{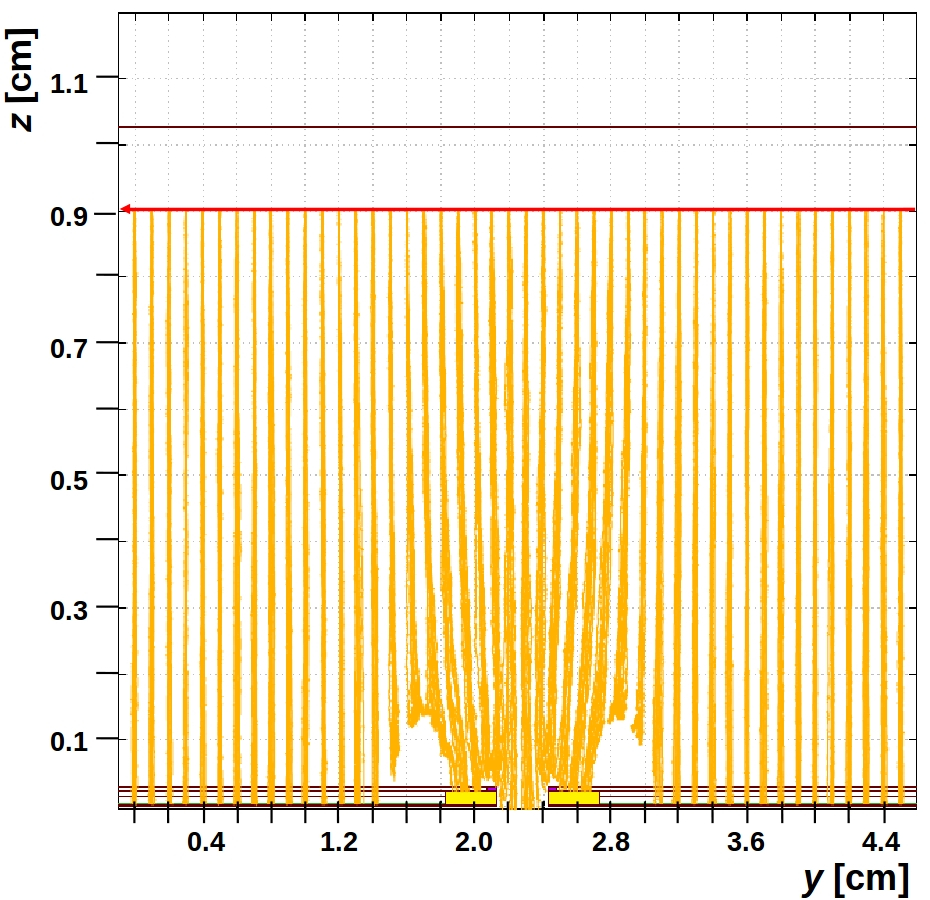}}
\subfigure[]
{\label{Drift-3}\includegraphics[height=0.3\textheight]{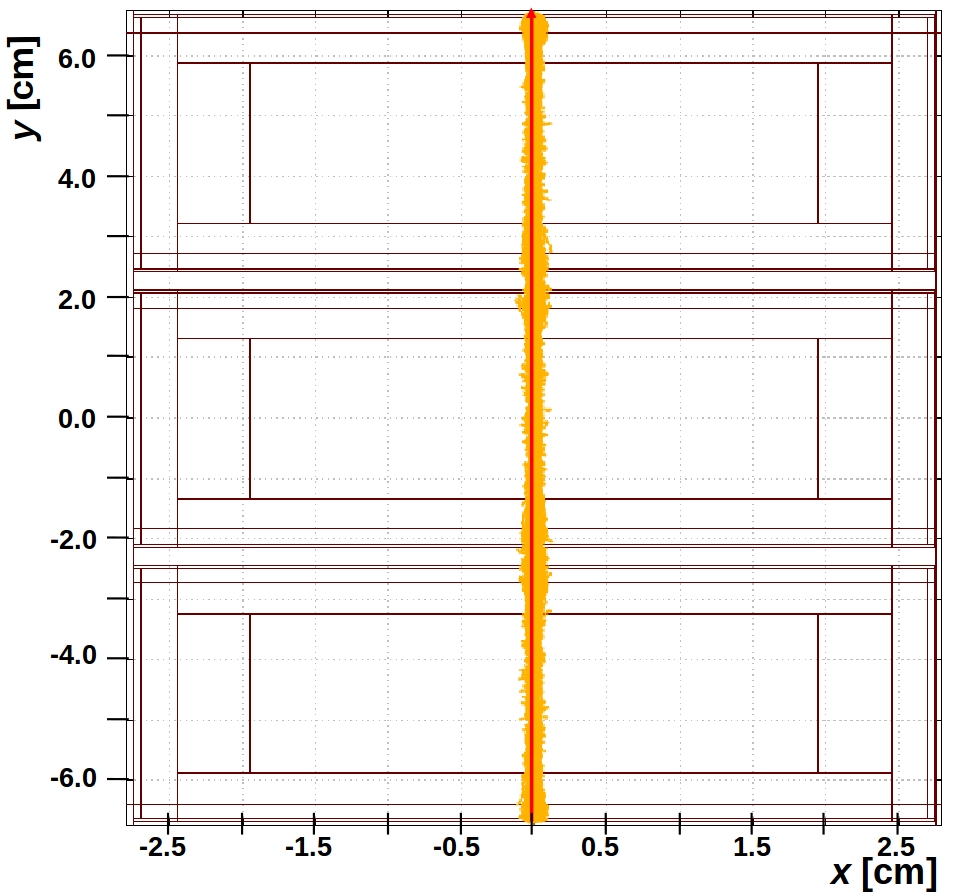}}
\subfigure[]
{\label{Drift-ZView-Case1}\includegraphics[height=0.3\textheight]{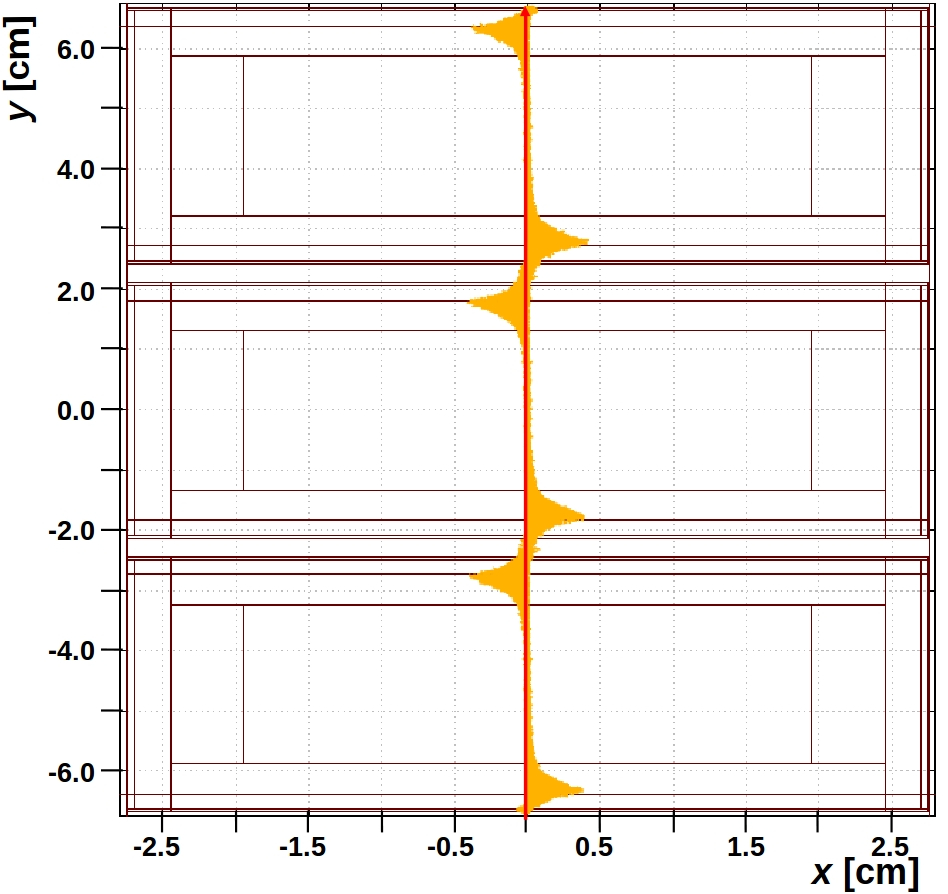}}
\caption{Drift of the electrons along the track are shown.
A small area adjacent to the photoresist is found to be free from any
hit while significant electron loss can be seen on the photoresist. (a) At
{\it{B}} = 0~T, the electrons are diffused. (b) At {\it{B}} = 1~T, diffusion is 
relatively small. The top view of the above two plots are shown in (c) and (d),
respectively.}
\label{Drift}
\end{figure}

\subsection{Drift Lines}
To study the effects of the field, a large number of electrons are released 
from the track as shown in Fig.~\ref{Area} at a distance of 
$5~\mathrm{mm}$ above the module.
These electrons drift towards the readout plane 
(Fig.~\ref{Drift}).
Near the edge of the module, the drift lines get distorted significantly. 
It is observed from Fig.~\ref{Drift-1} that $43\%$ of electrons are 
lost on the additional ground and the photoresist. 
From the drift lines, it is also clear that the number of electrons 
at the readout pads close to the module edge is much less in comparison to 
that at the central part of the module. 
The presence of the $\mathrm{1~T}$ magnetic field reduces the diffusion as 
shown in Fig.~\ref{Drift-YView-Case1}, as expected, but at the same 
time it introduces movement along {\it{y}} (for this particular configuration). 
This is natural since, under the application of magnetic field, the 
$\vec{{E}} \times \vec{{B}}$ force comes into play and 
modifies the drift lines. 
A view from the top, shown in Figs.~\ref{Drift-3} and ~\ref{Drift-ZView-Case1} 
complements the elevation projections presented in Figs.~\ref{Drift-1} and 
\ref{Drift-YView-Case1}. 
From these two views it is clear that the electrons have an inclination to 
drift towards the photoresist support or the nearby ground plane.
It may also be noted that, for the $\mathrm{1~T}$ magnetic field, the effect 
of diffusion is less and track distortion is restricted to a smaller 
fraction of the considered track. Although Fig.~\ref{Drift} present the 
situation corresponding to a given number of events along few tracks 
(10 tracks, 80 equidistant electrons per track) of what is clearly a 
statistical process, the indications are reasonably clear.

\begin{figure}[hbt]
\centering
\includegraphics[scale=0.5]{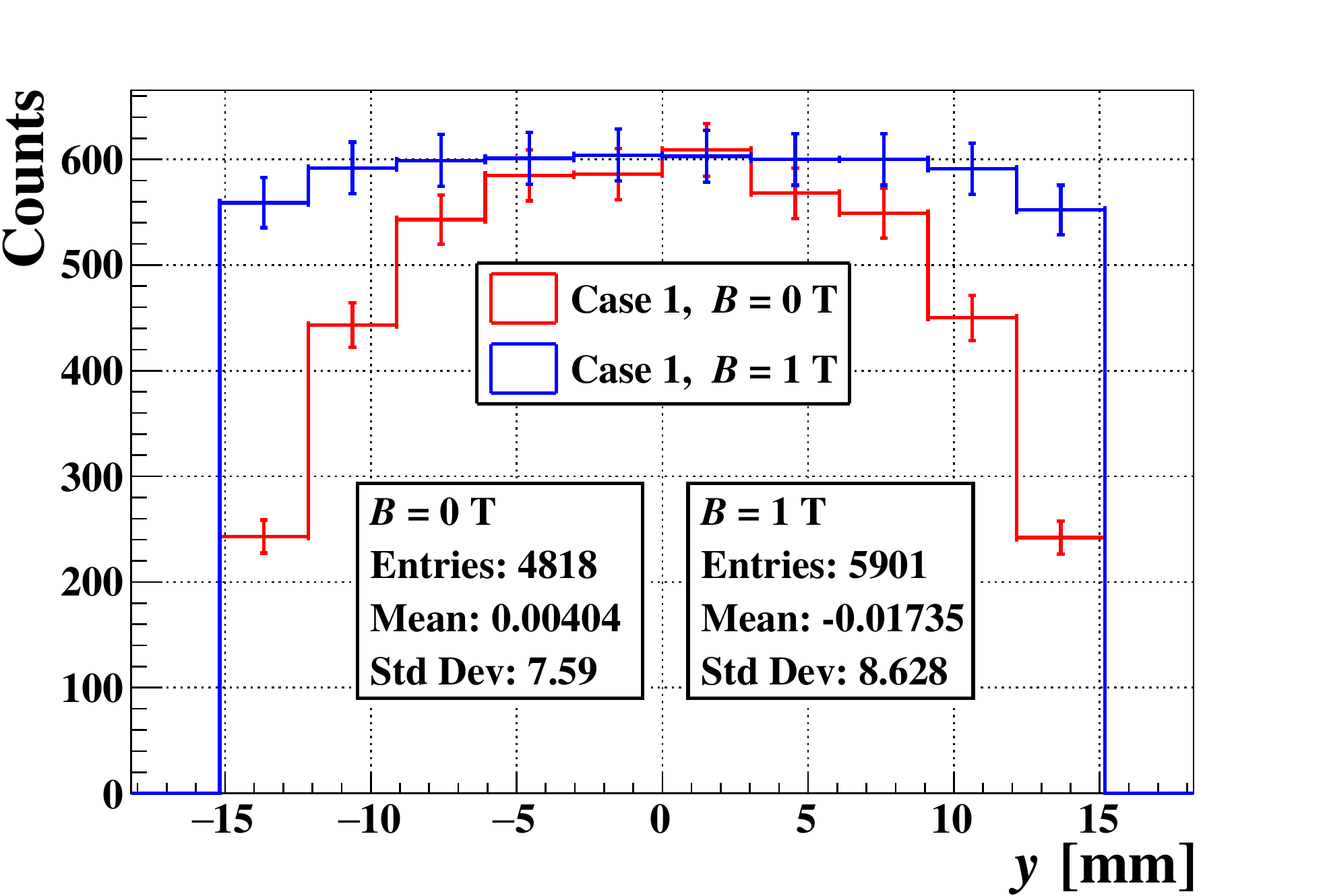}
\caption{Variation of counts along {\it{y}}.}
\label{PadRowVsCount}
\end{figure}

From the drift lines shown above, it is clear that a loss of efficiency will 
occur for pads close to the module edges.
In order to assess the loss, the {\it{y}} axis has has been 
divided into $3~\mathrm{mm}$ bins and the number of electrons within each 
bin has been counted. 
The variation of the count along the track has been presented in 
Fig.~\ref{PadRowVsCount}.  
The loss of efficiency, close to the edge is higher for ${\it{B}}~\mathrm{=0~T}$ 
(loss of more than $30\%$ up to $\mathrm{1~cm}$ from the module edge) than
that of ${\it{B}}~\mathrm{=1~T}$ (loss of more than $30\%$ up to $\mathrm{0.5~cm}$ 
from the module edge).

\subsection{Residual calculation}

In the experiment, the track is reconstructed out of the measured hit positions.
The residual of a pad hit is given by:
\begin{align}
\Delta {\it{x}} = {\it{x}}_{\mathrm{hit}} - {\it{x}}_{\mathrm{track}}    
\end{align}
where, ${\it{x}}_{\mathrm{hit}}$ is the true position of the hit and 
${\it{x}}_{\mathrm{track}}$ is the
estimated hit position based on the track fitting with the rest of the points.
The residuals provide an estimate of the distortion introduced during track
reconstruction.
In the simulation, as the track (i.e. the starting coordinates of the
electrons) is well defined, we can simplify the residual calculation without
track fitting.
The ending coordinates of the electrons (i.e. the final pad hit point) are
obtained from Garfield.
The residuals of the hits on the anode plane are estimated by taking the
difference between the respective start and end coordinates of the electrons.
They are then averaged over the number of events (tracks).

\begin{figure}[hbt]
\centering
\subfigure[]
{\label{HistogramX}\includegraphics[scale=0.375]{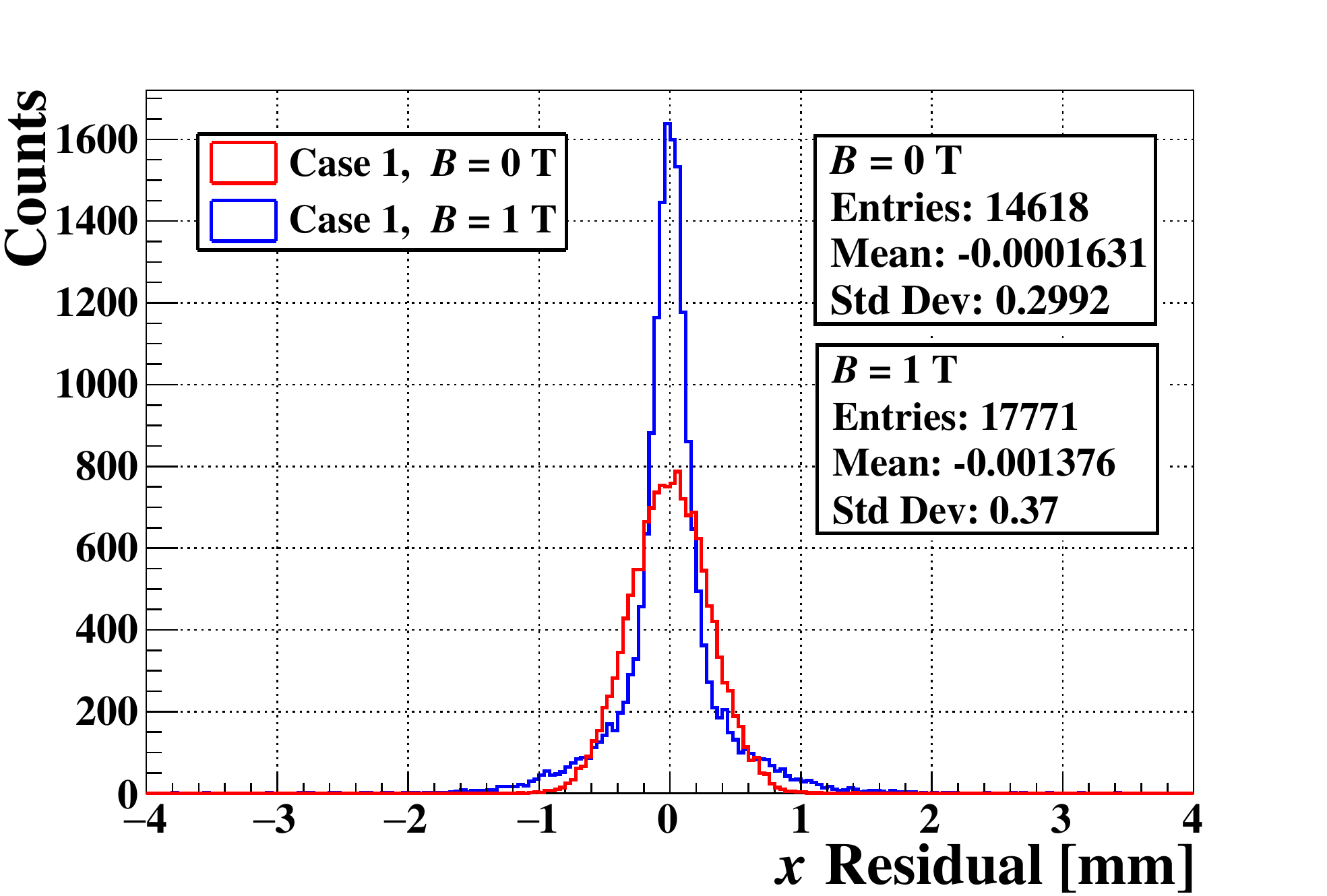}}
\subfigure[]
{\label{HistogramY}\includegraphics[scale=0.375]{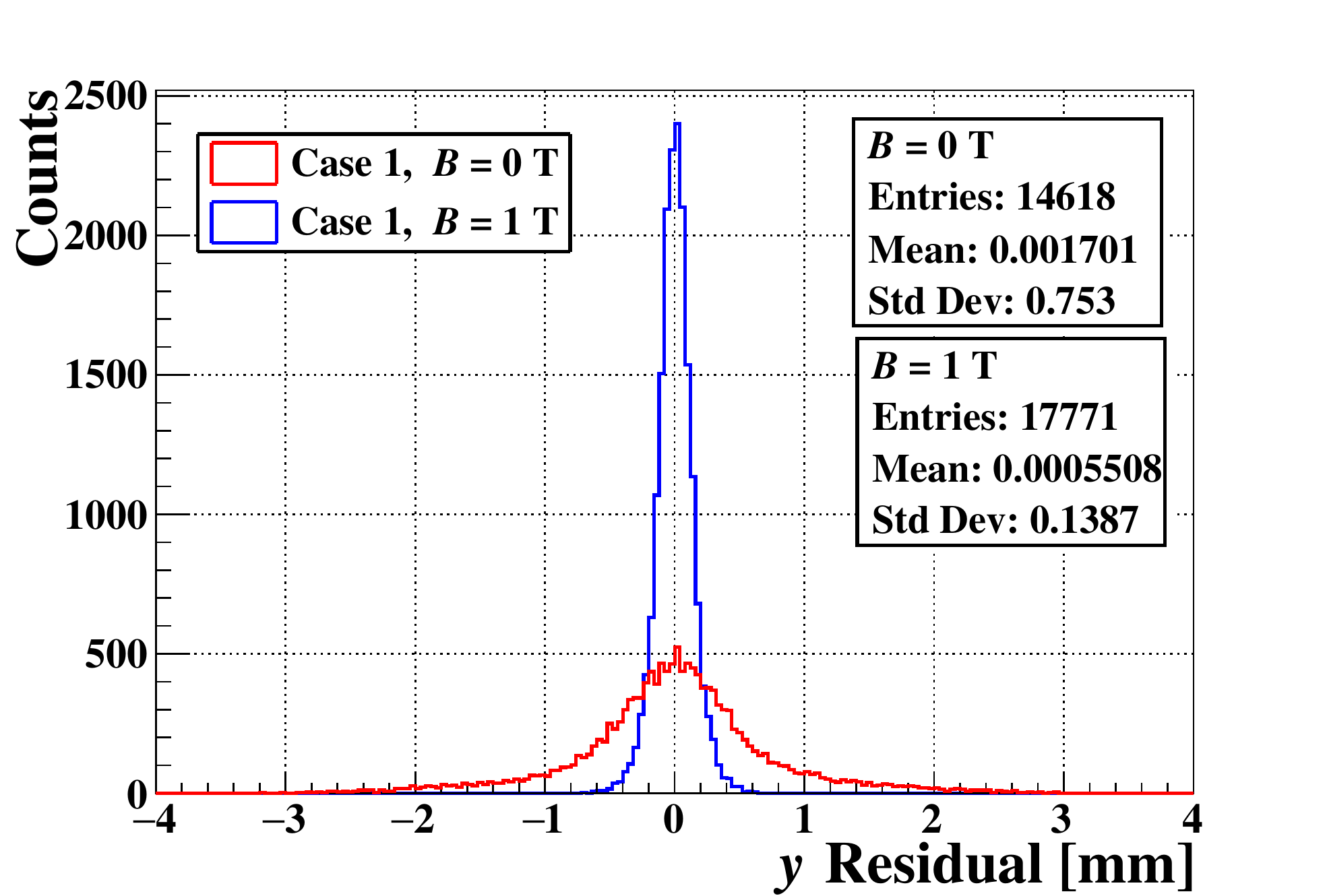}}
\caption{Histogram of (a) {\it{x}} and (b) {\it{y}} residual in ${\it{B}}= 0~\mathrm{T}$ and ${\it{B}} = 1~\mathrm{T}$.}
\label{Histogram}
\end{figure}

\begin{figure}[hbt]
\centering
\subfigure[]
{\label{ResidueX}\includegraphics[scale=0.375]{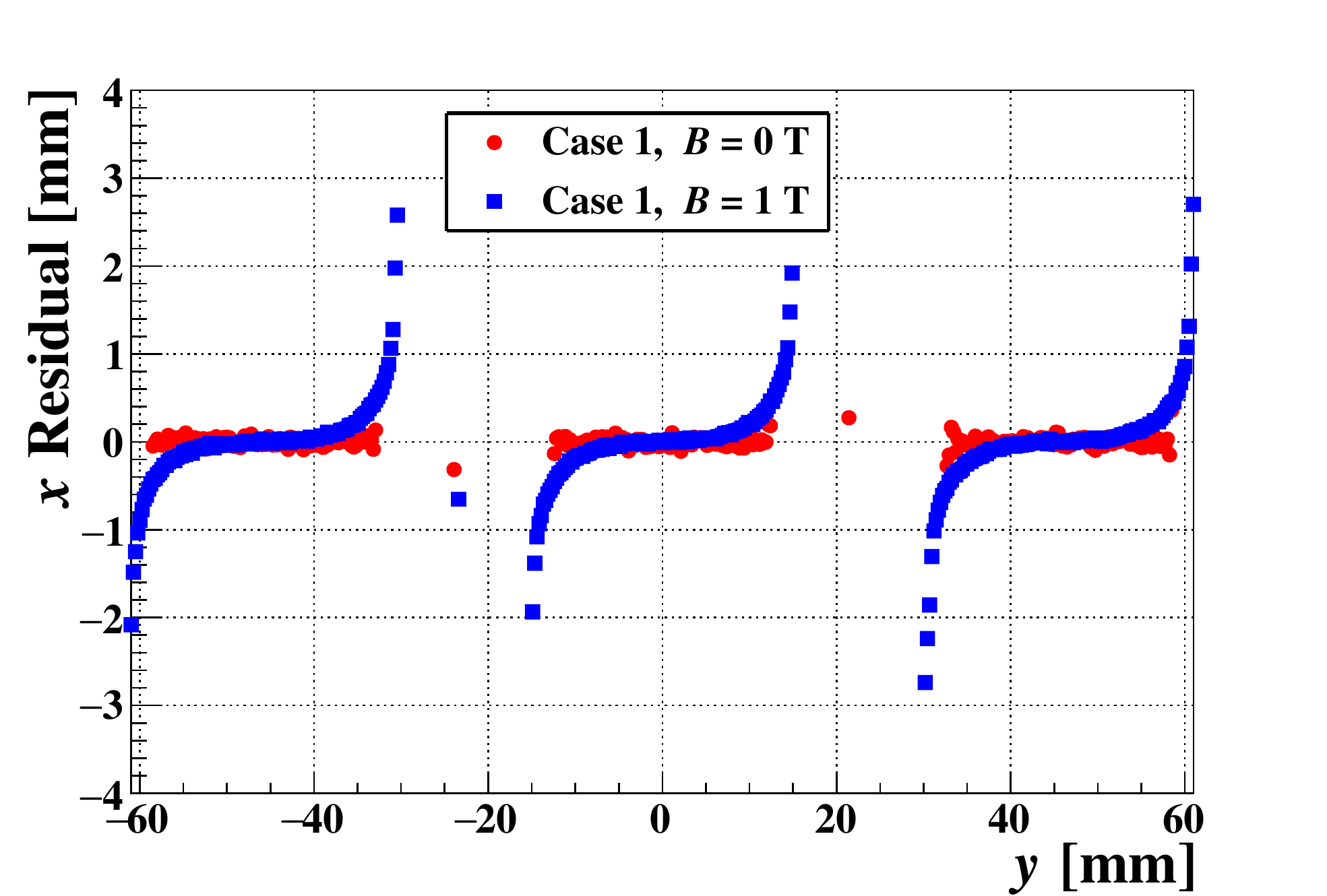}}
\subfigure[]
{\label{ResidueY}\includegraphics[scale=0.375]{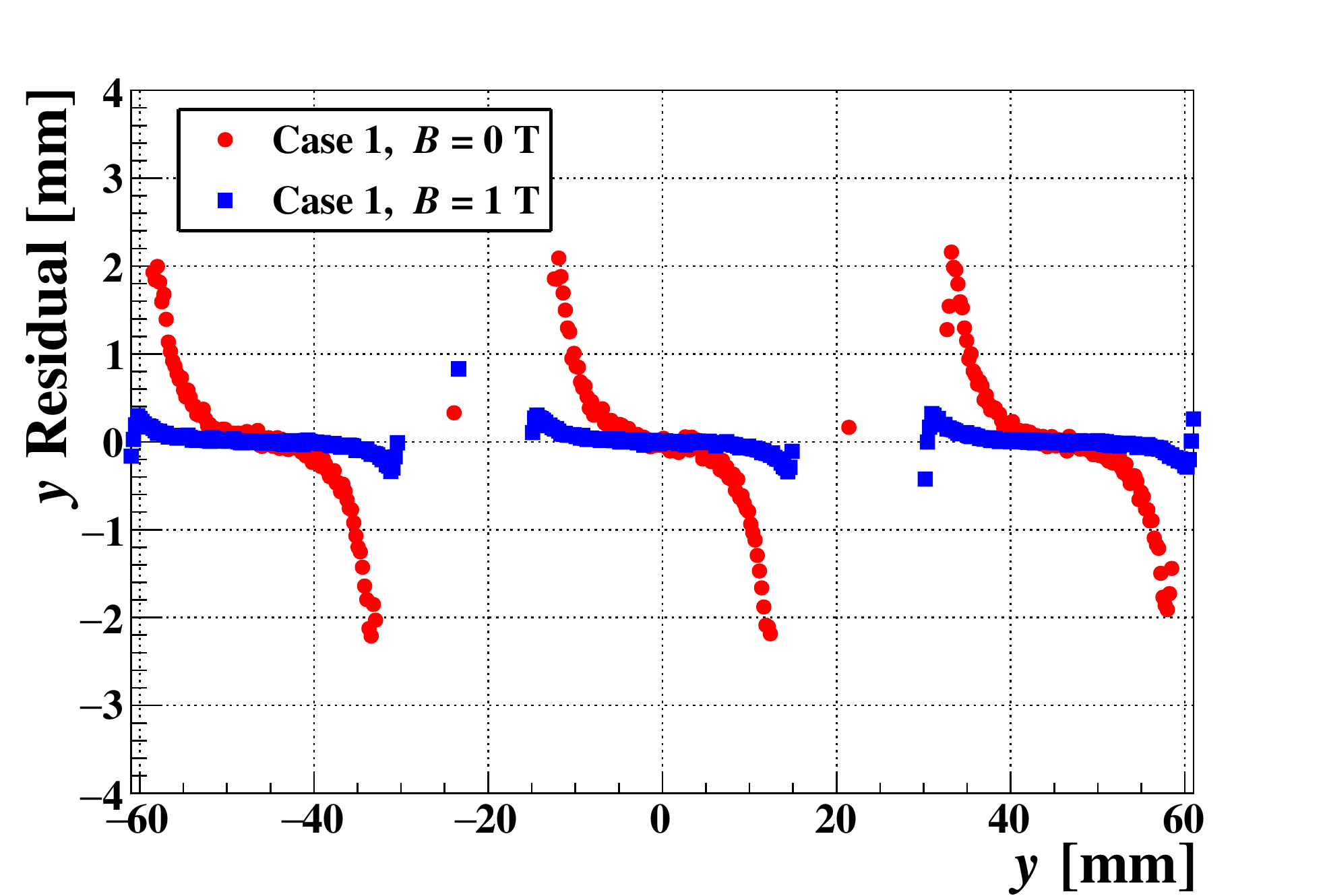}}
\caption{ (a) {\it{x}} and (b) {\it{y}} residual in ${\it{B}}= 0~\mathrm{T}$ and ${\it{B}} = 1~\mathrm{T}$.}
\label{Residue}
\end{figure}

From the residual histograms (Fig.~\ref{Histogram}), it can be observed that
residual along {\it{x}} reduce significantly in the middle of the module
due to the presence of a magnetic field.
This is easily explained by noting that magnetic field decreases diffusion
since it acts as an additional constraint that makes the charged particle
follow electric field lines.
The same trend is observed in the residual along {\it{y}}.
The {\it{x}}-residual plots without the magnetic field (Fig.~\ref{ResidueX})
does not show any distortion in the middle of the modules.
The shift between the modules is clearly visible.
When a magnetic field is applied, the magnetic
field together with the transverse components of the electric
field gives rise to the Lorentz force near the gap (Fig.~\ref{Drift}).
This force leads to a distortion along {\it{x}}
and as a result, the magnitude of the residuals are high near the
module gaps, as shown in Fig.~\ref{ResidueX}.

The {\it{y}}-residual plots without the magnetic field shows a 
large distortion near the edge due to the electrical field inhomogeneity.
In presence of the magnetic field, the lower diffusion helps to reduce the
the residual value (Figs.~\ref{HistogramY} and ~\ref{ResidueY}).

\begin{figure}[hbt]
\centering
\subfigure[]
{\label{SpatialX}\includegraphics[scale=0.375]{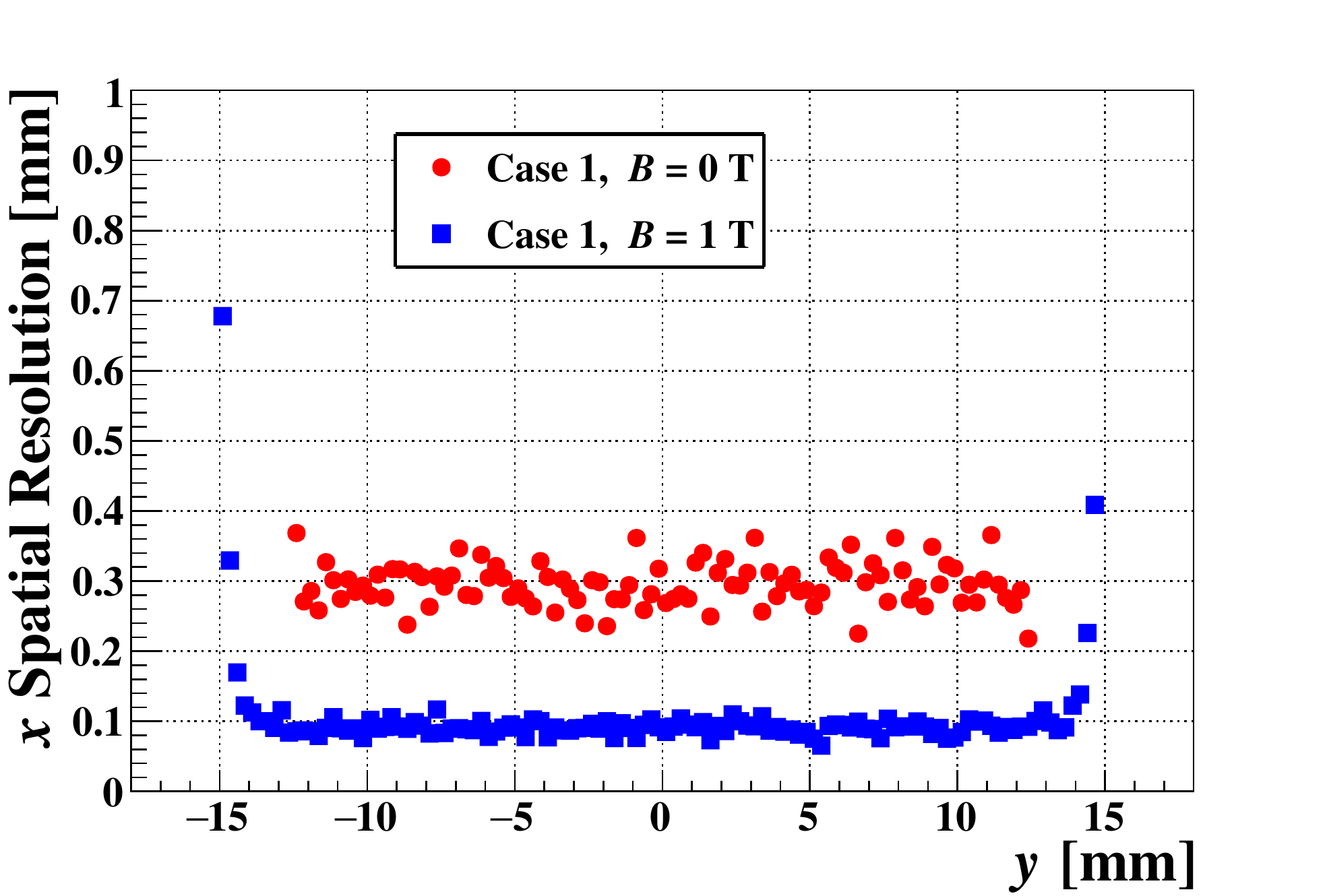}}
\subfigure[]
{\label{SpatialY}\includegraphics[scale=0.375]{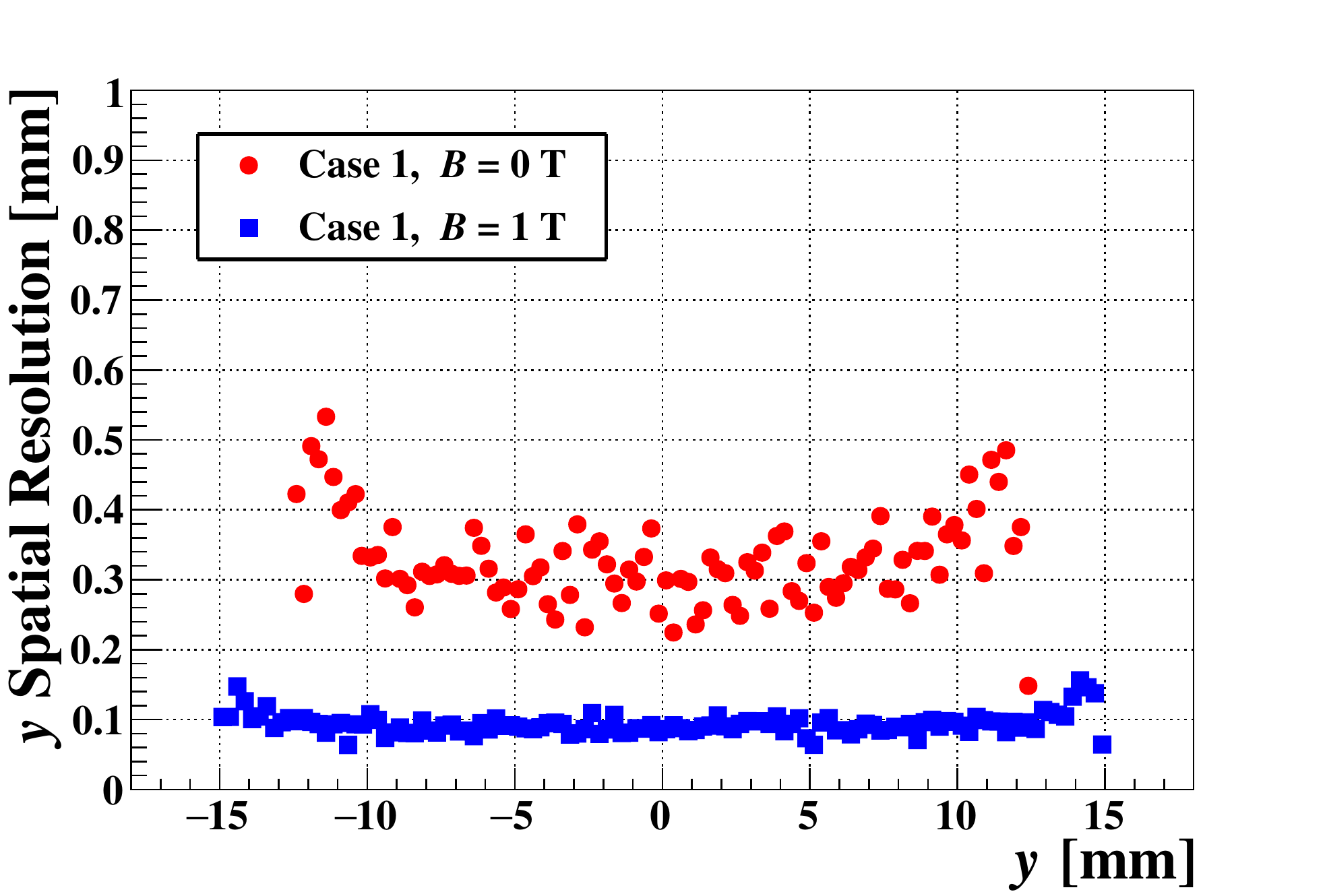}}
\caption{(a) {\it{x}} and (b) {\it{y}} spatial resolution in ${\it{B}}= 0~\mathrm{T}$ and ${\it{B}} = 1~\mathrm{T}$.}
\label{Spatial}
\end{figure}

Although the numerical model is simpler and smaller than the real detector, we have compared the numerical estimates against the experimentally 
measured values, as shown in fig.~\ref{ExptResidue}.
The length of the numerical module is only 
$\mathrm{5.5}\times\mathrm{4.25~cm^2}$, in comparison to 
$\mathrm{22}\times\mathrm{17~cm^2}$ of the real module. 
As a result, the {\it{y}}-range in the numerical estimate ($\mathrm{3.644~cm}$) is 
much smaller in comparison to the experimental one ($\mathrm{16.394~cm}$). 
By comparing Figs.~\ref{Expt1} and ~\ref{ResidueX}, it is observed 
that for the ${\it{B}}~\mathrm{= 0~T}$ case, distortion in both the cases are 
found to be around $\mathrm{0.5~mm}$, while for the ${\it{B}}~\mathrm{ = 1~T}$ case,
the distortions are around $\mathrm{2~mm}$. 
Thus, despite the use of a smaller and simpler model for the numerical 
simulation, the estimates are qualitatively and quantitatively comparable 
to the experimental results.  
It may be safely concluded that the numerical model represents the 
experimental scenario reasonably well and may be used for further studies 
on possible design modifications.
The {\it{x}} and {\it{y}} spatial resolutions are plotted in Fig.~\ref{Spatial}. 
Spatial resolution is worsened near the edges. 
In presence of the magnetic field, the resolution improves.  

\subsection{Possible design modification}

\begin{figure}[hbt]
\centering
\includegraphics[scale=0.8]{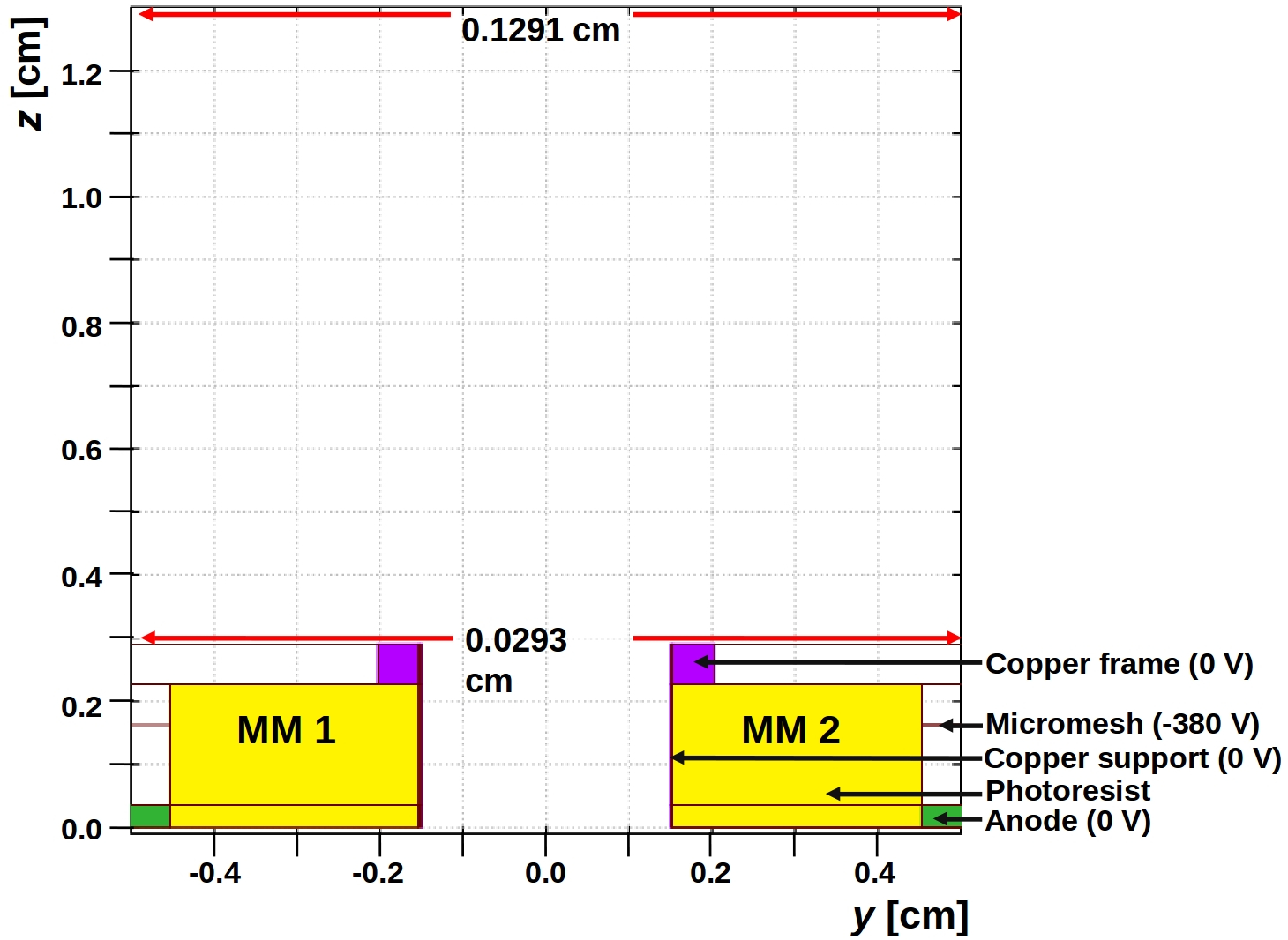}
\caption{Side view of different numerical modules. The anode and copper frame are no longer continuous and a photoresist separates the two.}
\label{ModifiedGeometry}
\end{figure}

Since the electrostatic field inhomogeneity has been identified as the
principal deciding factor determining the distortion during track
reconstruction in a multi-module TPC, the obvious approach to reduce the
distortion is to try and modify the electrostatic field configuration near
the module edges.
The main difference between the proposed configurations and the original one
is the attempt to maintain the copper frame and micro-mesh at the same
potential so that the possibility of generating a transverse electric field is
minimized.
Since the anode and the copper frame no longer has the same potential, they
are no longer continuous and a photoresist separates the two to ensure
electrical insulation (Fig.~\ref{ModifiedGeometry}).
Different possible configurations which have been considered, are listed in
table~\ref{table2}.

\begin{table}[hbt]
\centering
\begin{tabular}{|c|c|c|c|c|}
\hline
Case & Description & Mesh & Copper Frame & Anode \\
& & Voltage & Voltage & Voltage \\
\hline
Case 1 & Anode and copper  & -380 V & 0 V & 0 V \\
& frame in contact & & & \\
\hline
Case 2 & Photoresist between & -380 V & 0 V & 0 V \\
\cline{1-1}\cline{3-5}
Case 3 & anode and copper & 0 V & 0 V & 380 V \\
\cline{1-1}\cline{3-5}
Case 4 & frame & -380 V & -380 V & 0 V \\
\hline
\end{tabular}
\caption{Different possible design modification of MM modules}
\label{table2}
\end{table}

\begin{figure}[hbt]
\centering
\subfigure[]
{\label{Study1-EY-Run2}\includegraphics[scale=0.375]{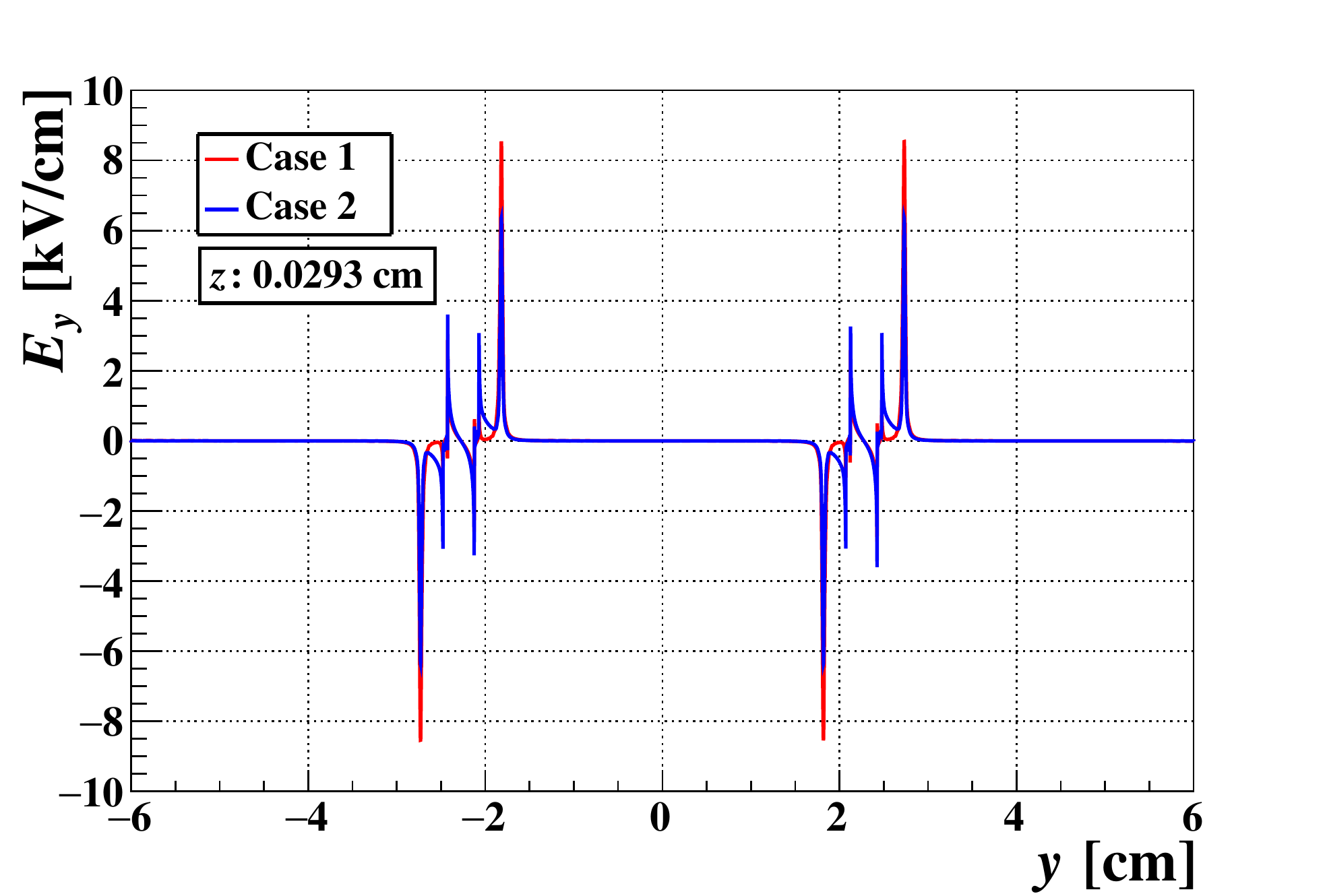}}
\subfigure[]
{\label{Study1-EY-Run3}\includegraphics[scale=0.375]{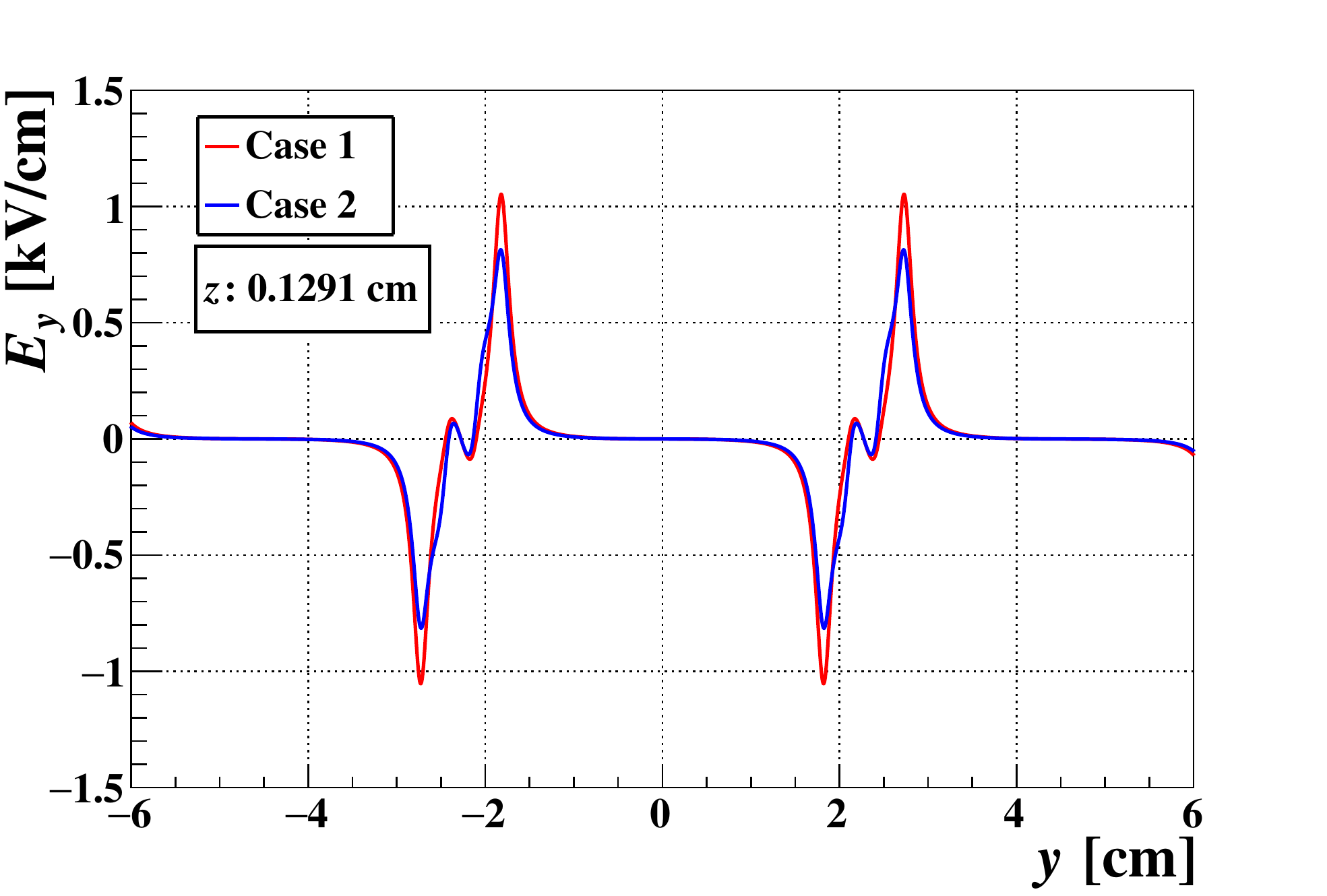}}
\subfigure[]
{\label{Study1-EZ-Run2}\includegraphics[scale=0.375]{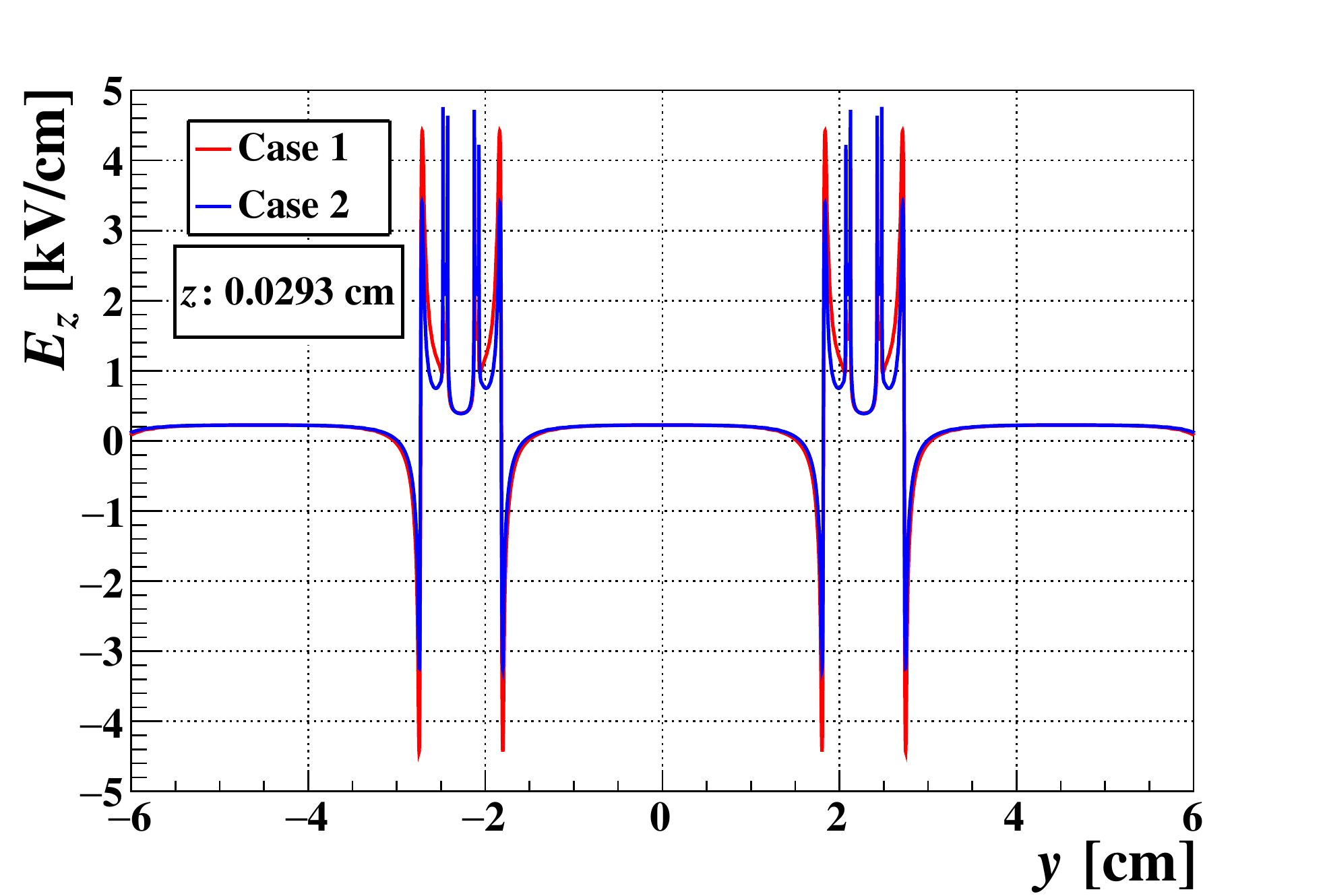}}
\subfigure[]
{\label{Study1-EZ-Run3}\includegraphics[scale=0.375]{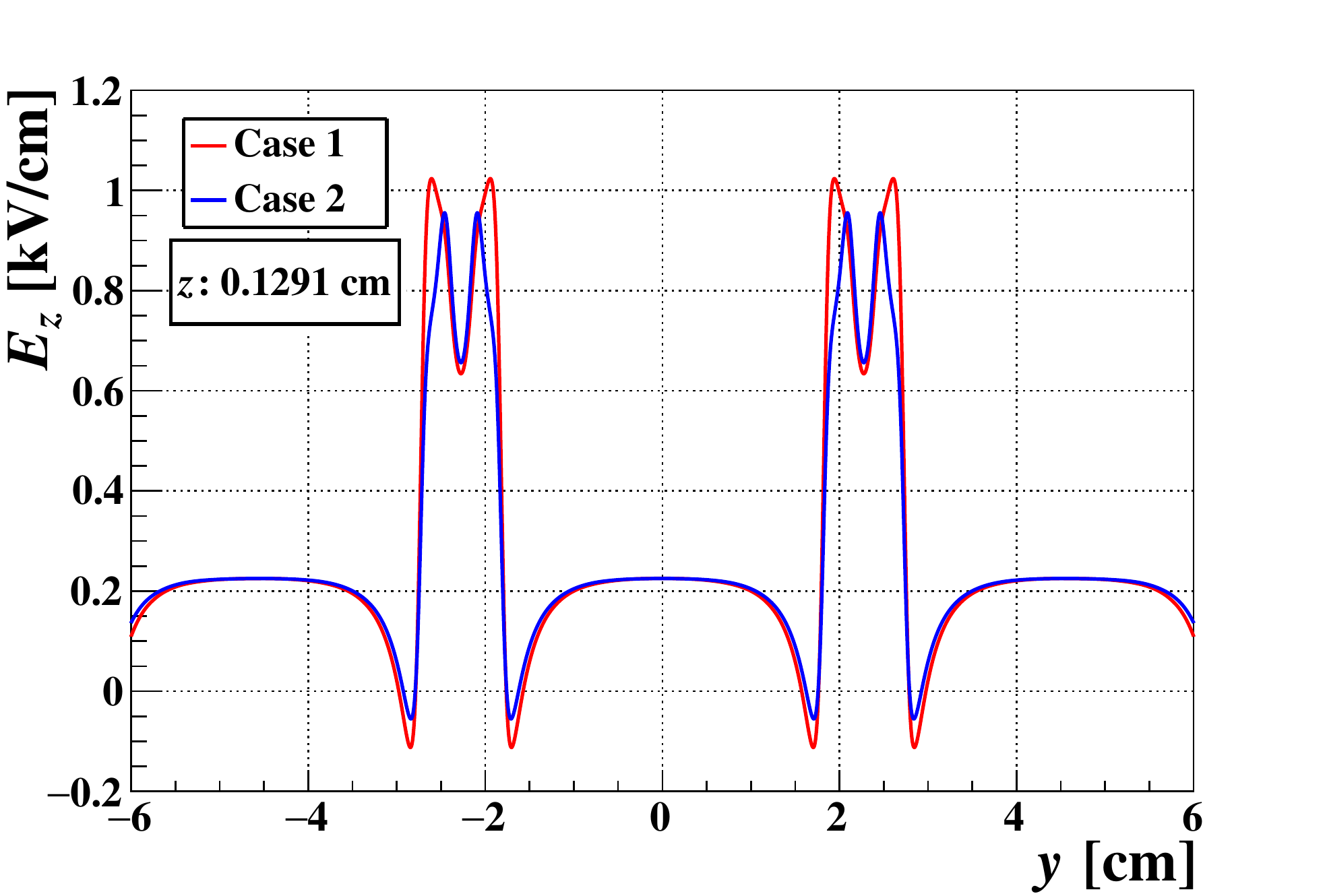}}
\caption{$\it{E}_{\it{y}}$ along (a) $\it{z} = \mathrm{0.0293}~\mathrm{cm}$ and (b) $\it{z} = \mathrm{0.1291}~\mathrm{cm}$, $\it{E}_{\it{z}}$ along (c) $\it{z} = \mathrm{0.0293}~\mathrm{cm}$ and (d) $\it{z} = \mathrm{0.1291}~\mathrm{cm}$ due to the inclusion of a photoresist separator.}
\label{Study1}
\end{figure}

\begin{figure}[hbt]
\centering
\subfigure[]
{\label{Study1-XHist}\includegraphics[scale=0.375]{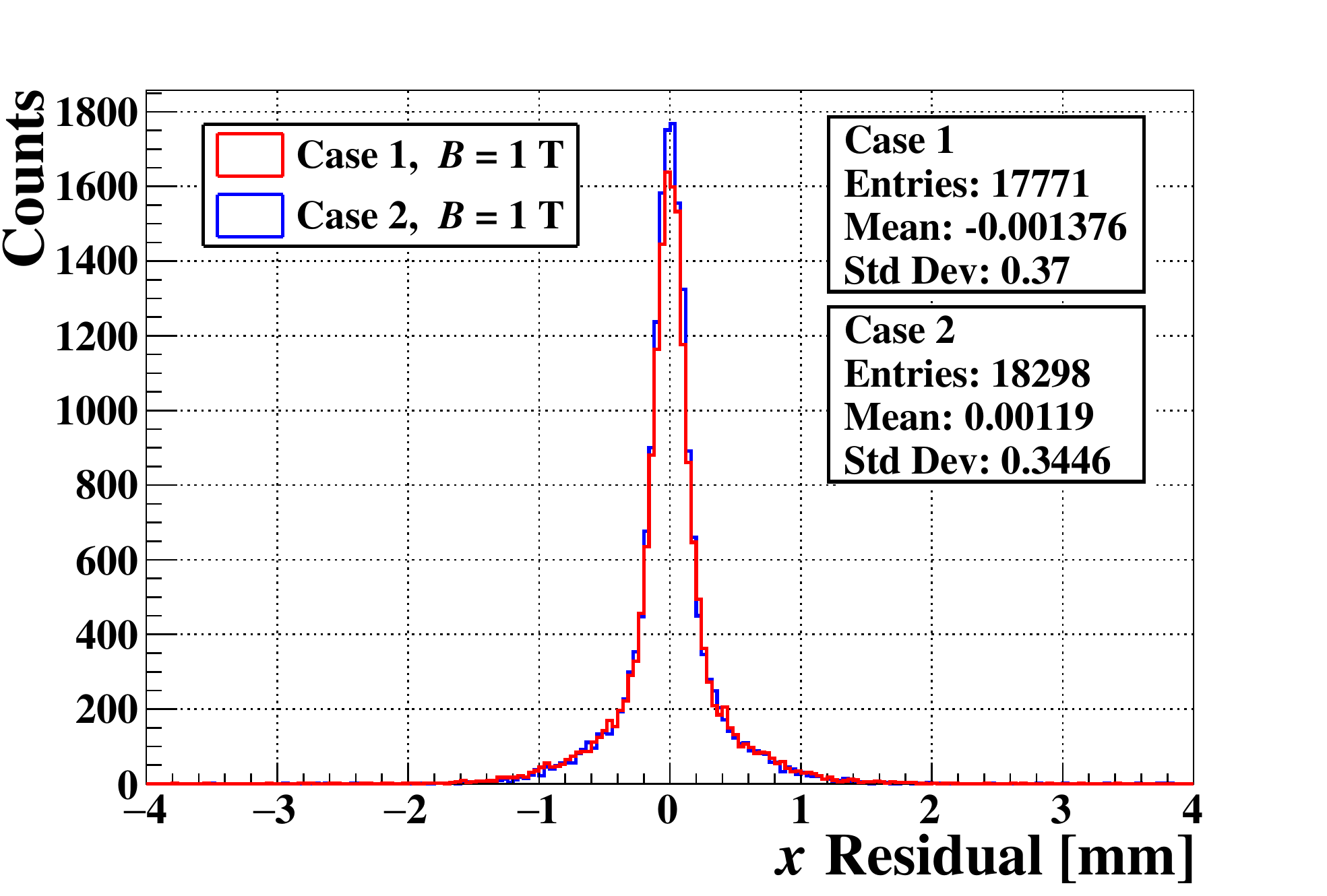}}
\subfigure[]
{\label{Study1-XRes}\includegraphics[scale=0.375]{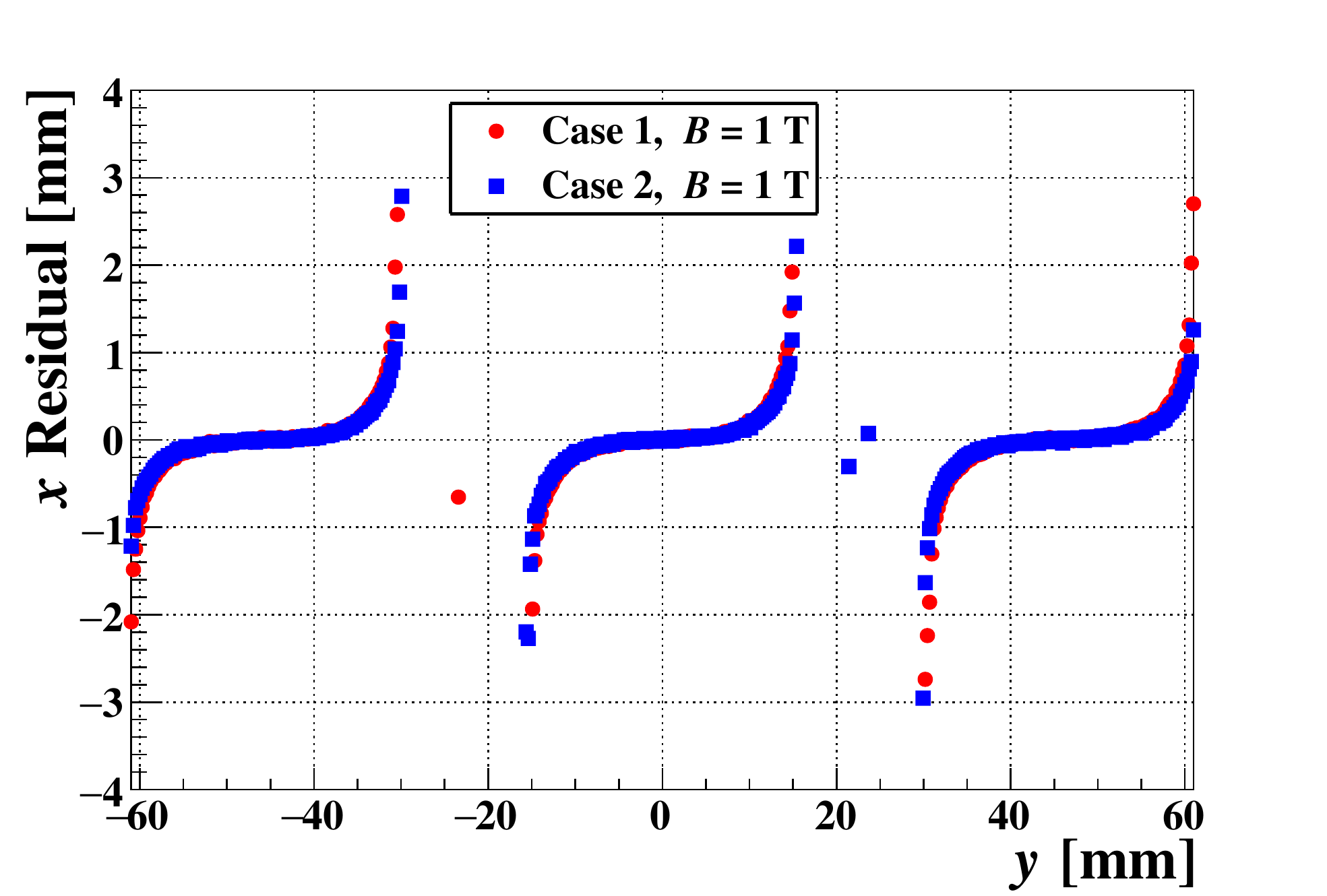}}
\subfigure[]
{\label{Study1-YHist}\includegraphics[scale=0.375]{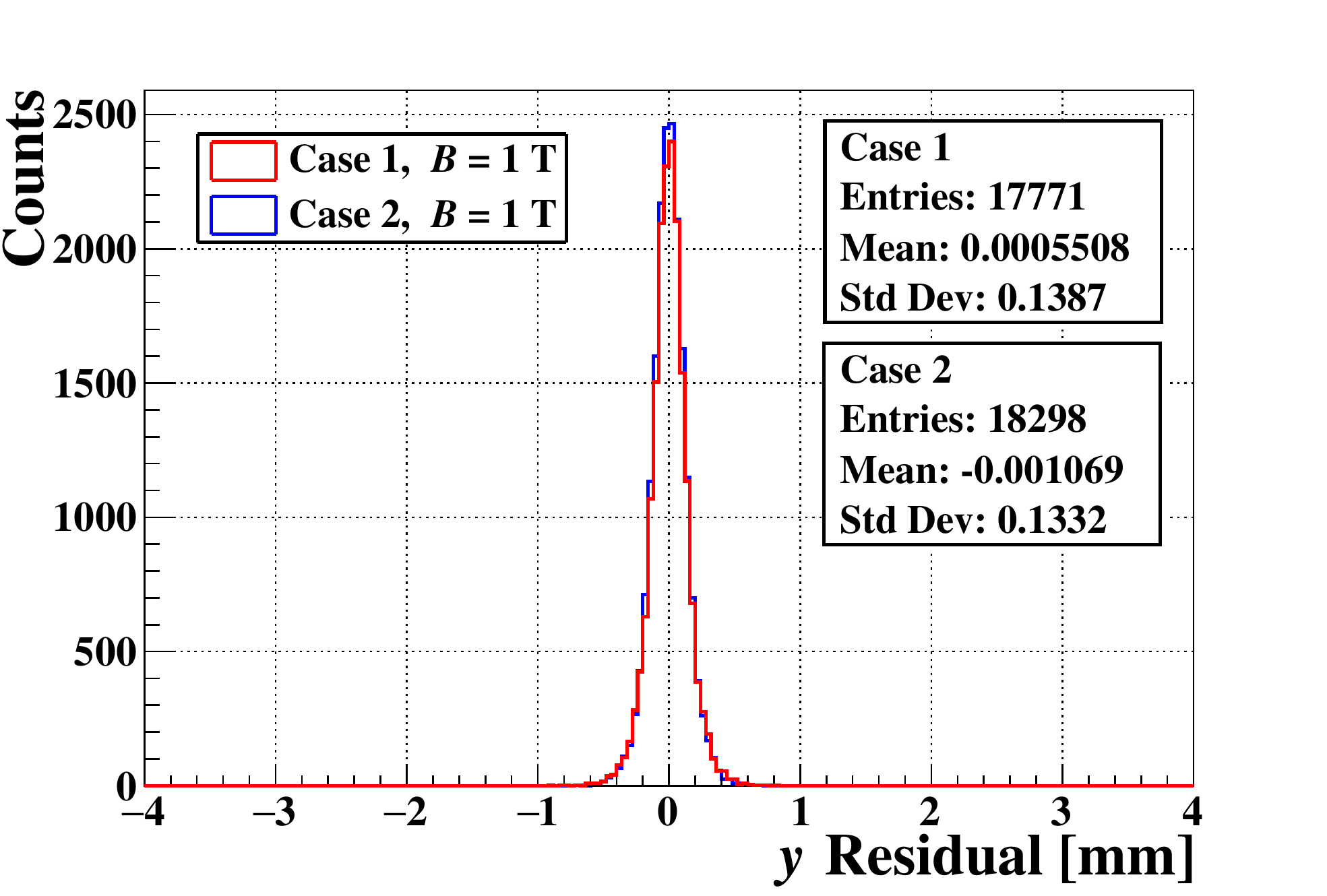}}
\subfigure[]
{\label{Study1-YRes}\includegraphics[scale=0.375]{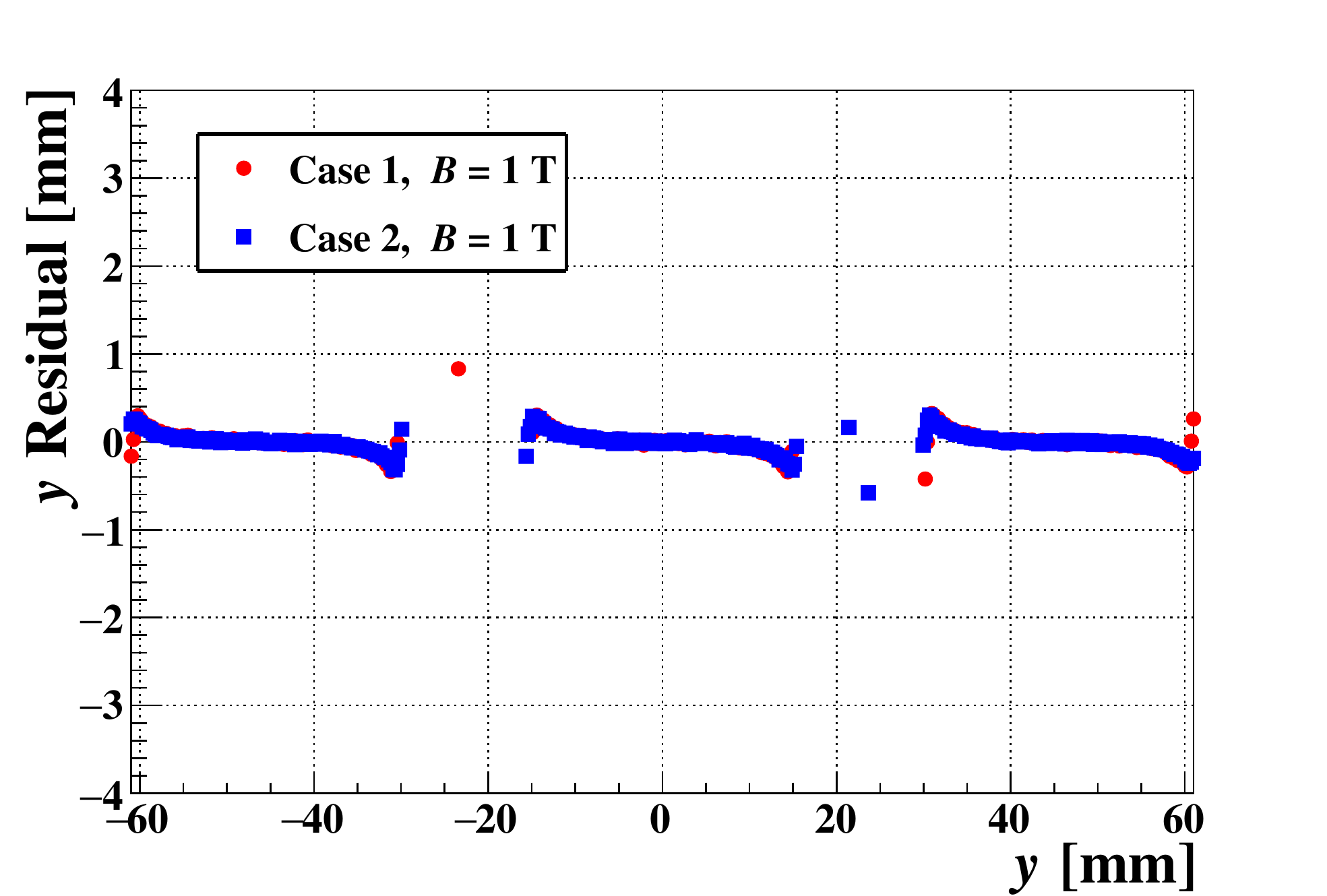}}
\caption{Comparison of (a) {\it{x}} (b) {\it{y}} residual histogram and comparison of (c) {\it{x}} (d) {\it{y}} residual due to the inclusion of a photoresist separator.}
\label{Study1Residue}
\end{figure}

\begin{figure}[hbt]
\centering
\subfigure[]
{\label{Study2-EY-Run2}\includegraphics[scale=0.375]{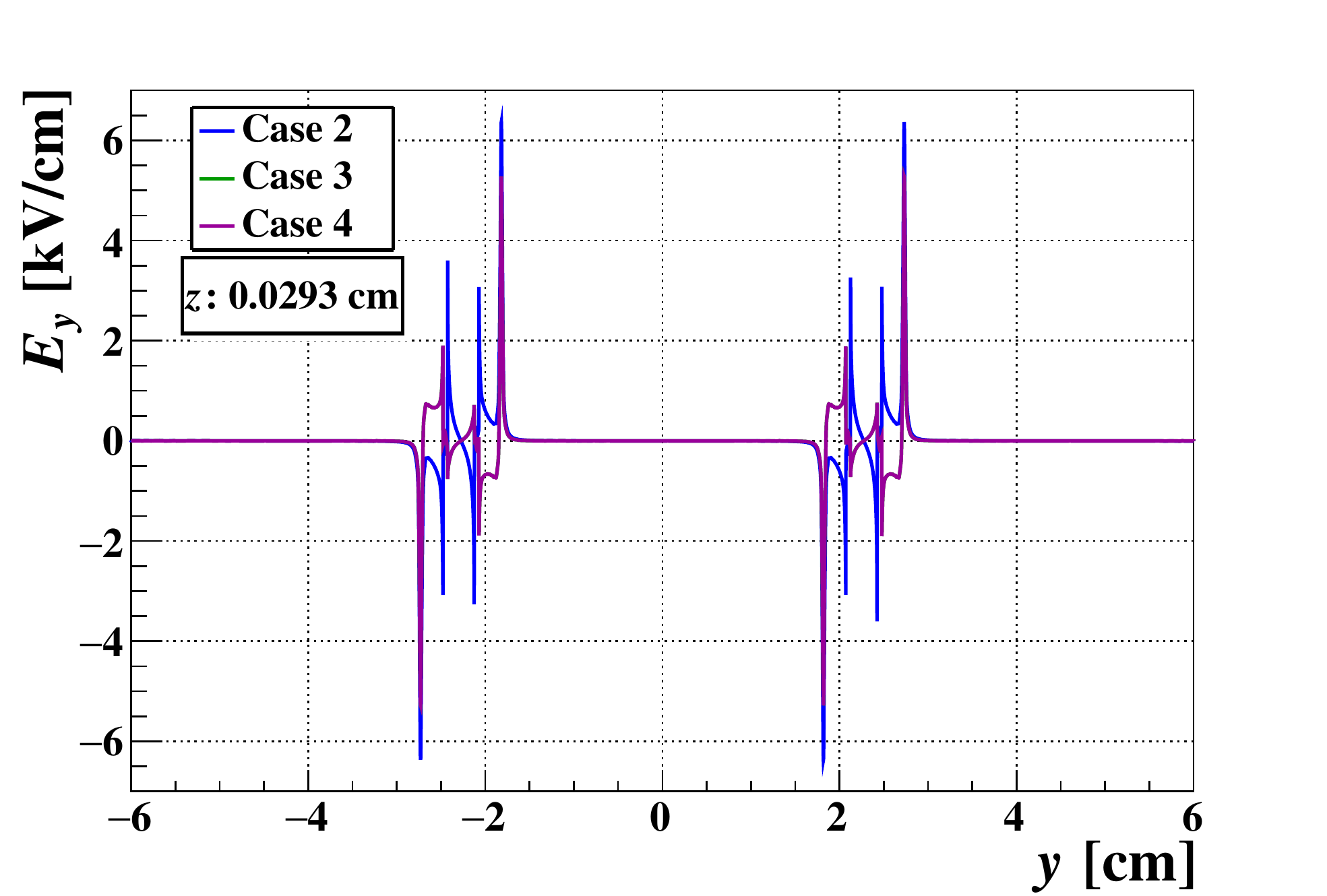}}
\subfigure[]
{\label{Study2-EY-Run3}\includegraphics[scale=0.375]{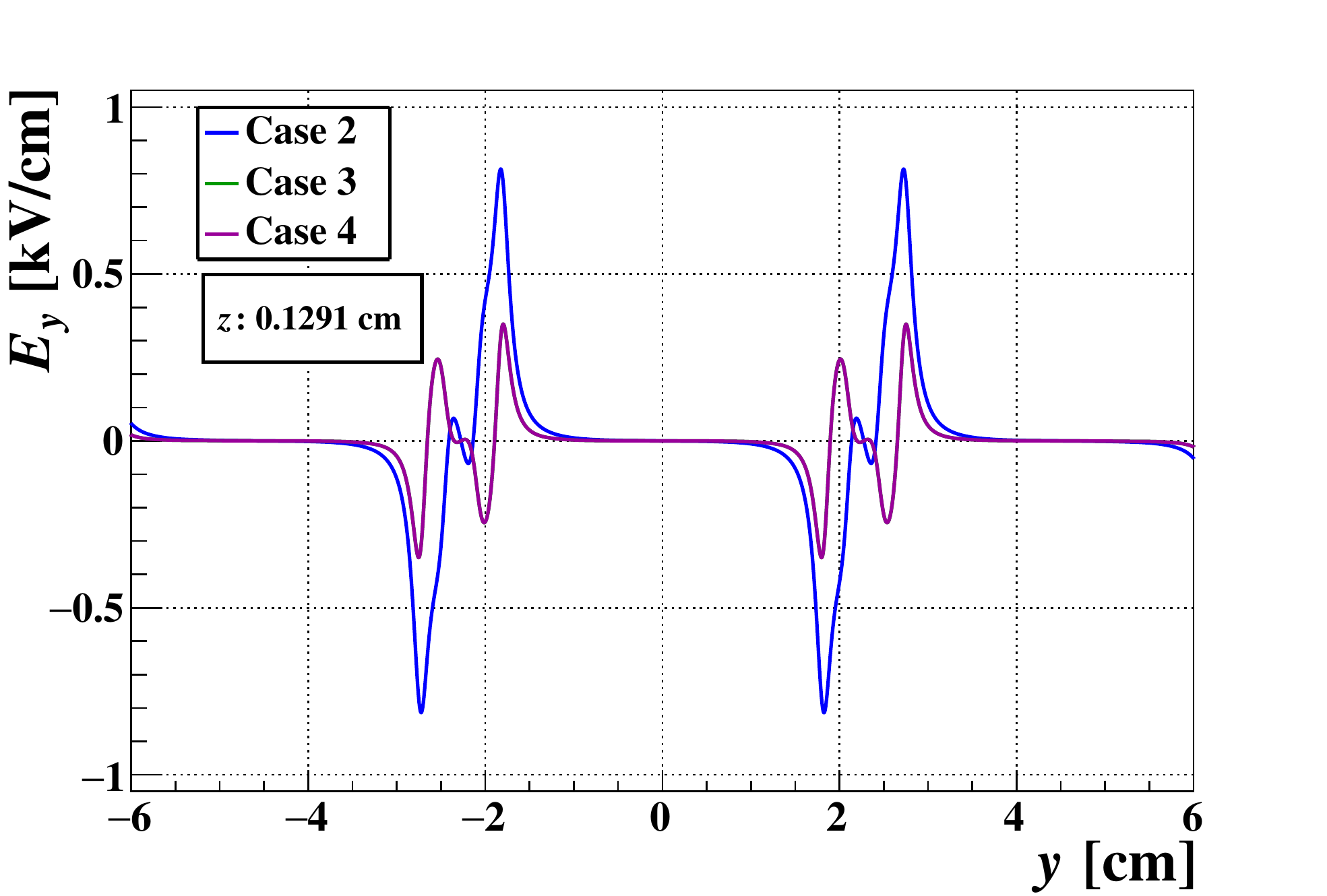}}
\subfigure[]
{\label{Study2-EZ-Run2}\includegraphics[scale=0.375]{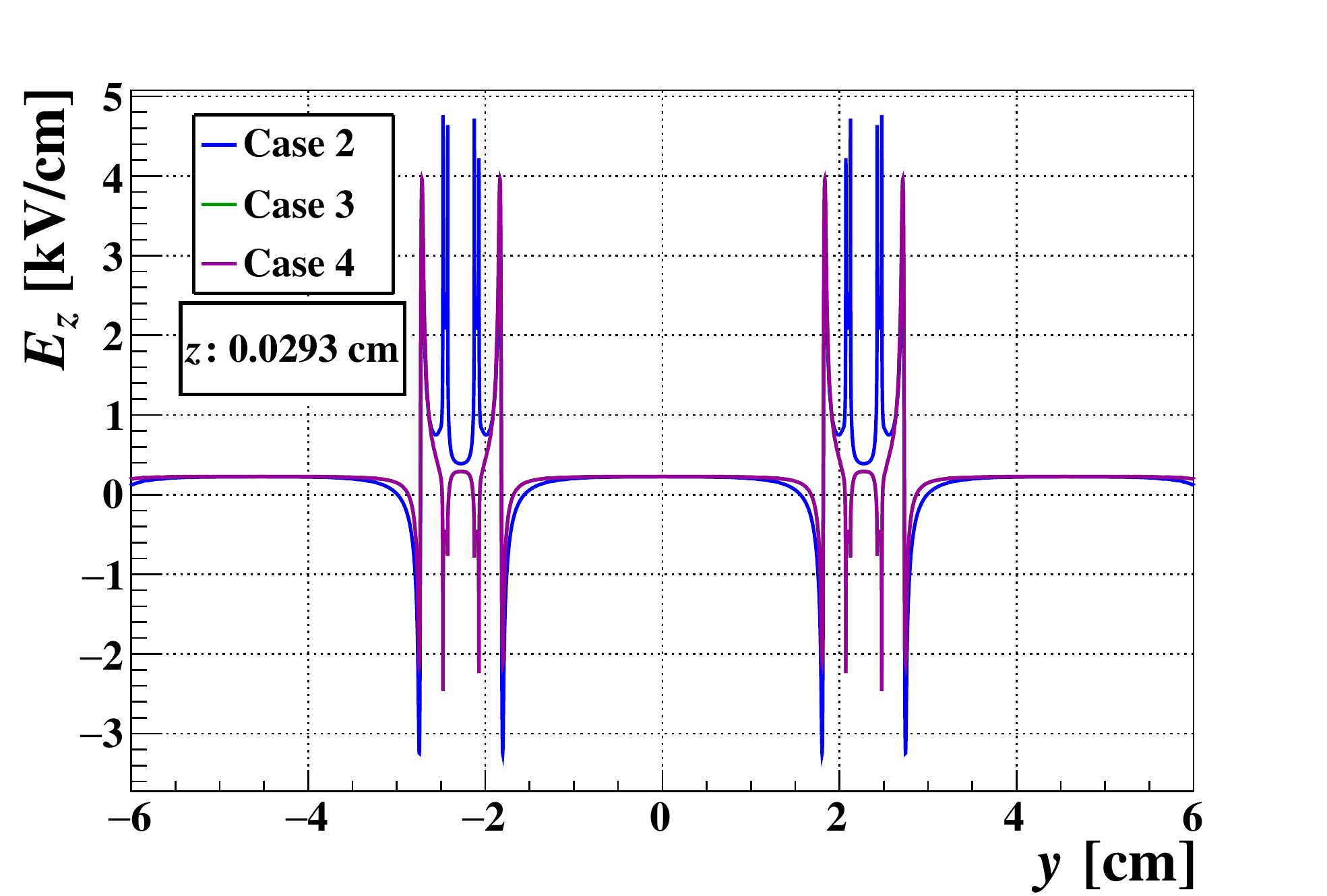}}
\subfigure[]
{\label{Study2-EZ-Run3}\includegraphics[scale=0.375]{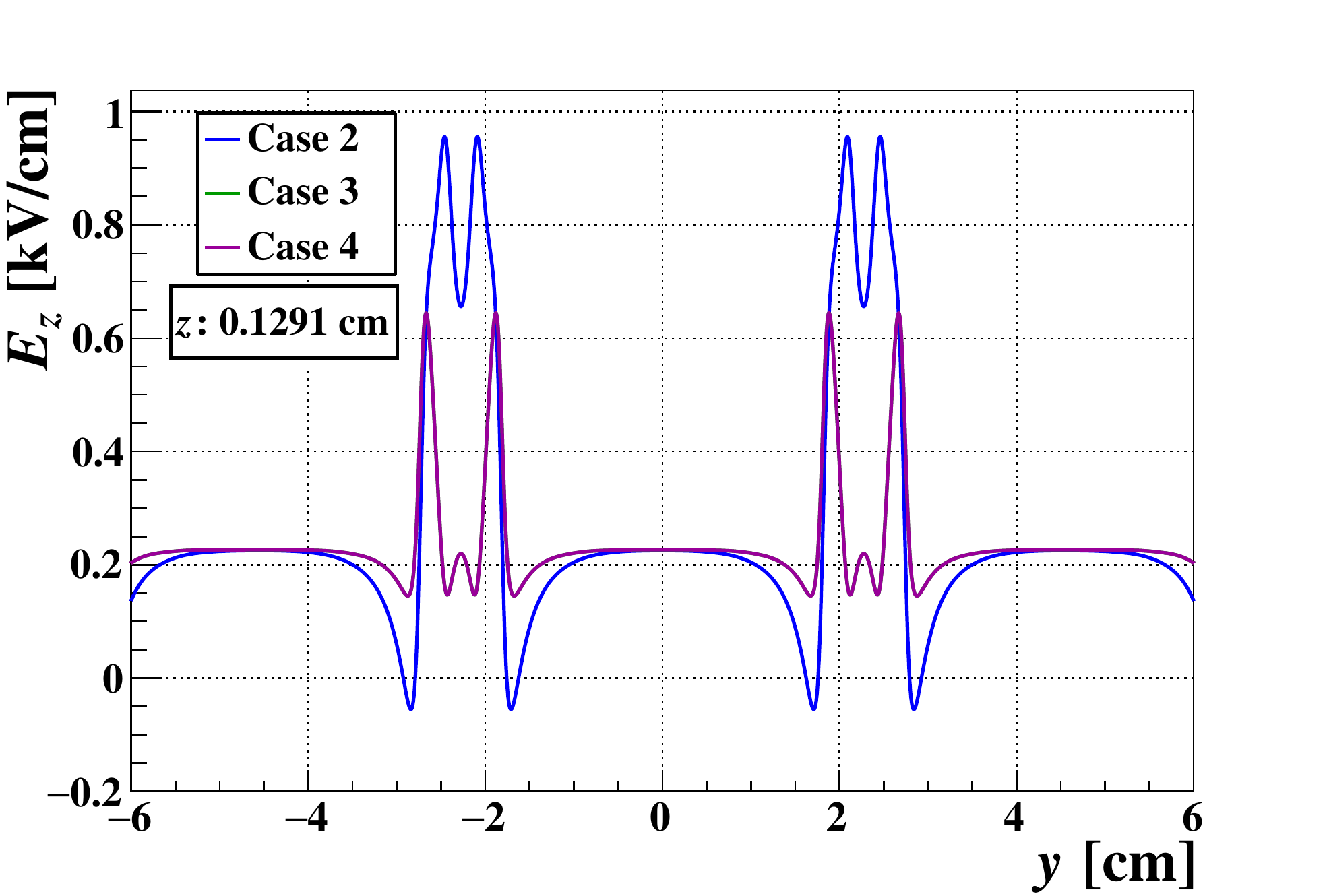}}
\caption{$\it{E}_{\it{y}}$ along (a) $\it{z} = \mathrm{0.0293}~\mathrm{cm}$ and (b) $\it{z} = \mathrm{0.1291}~\mathrm{cm}$, $\it{E}_{\it{z}}$ along (c) $\it{z} = \mathrm{0.0293}~\mathrm{cm}$ and (d) $\it{z} = \mathrm{0.1291}~\mathrm{cm}$ for various potential applied to the copper frame.}
\label{Study2}
\end{figure}

\begin{figure}[hbt]
\centering
\subfigure[]
{\label{Study2-XHist}\includegraphics[scale=0.375]{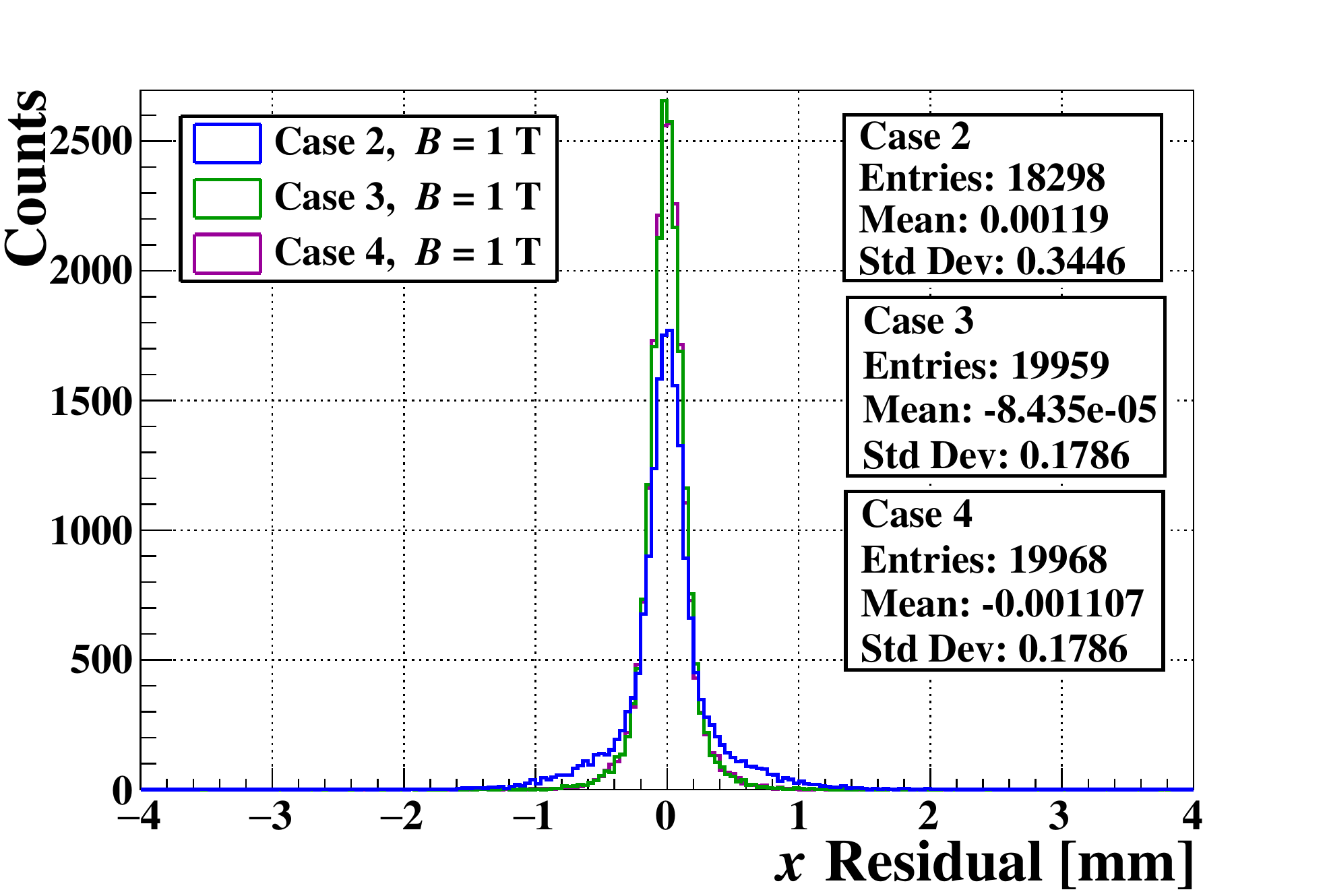}}
\subfigure[]
{\label{Study2-XRes}\includegraphics[scale=0.375]{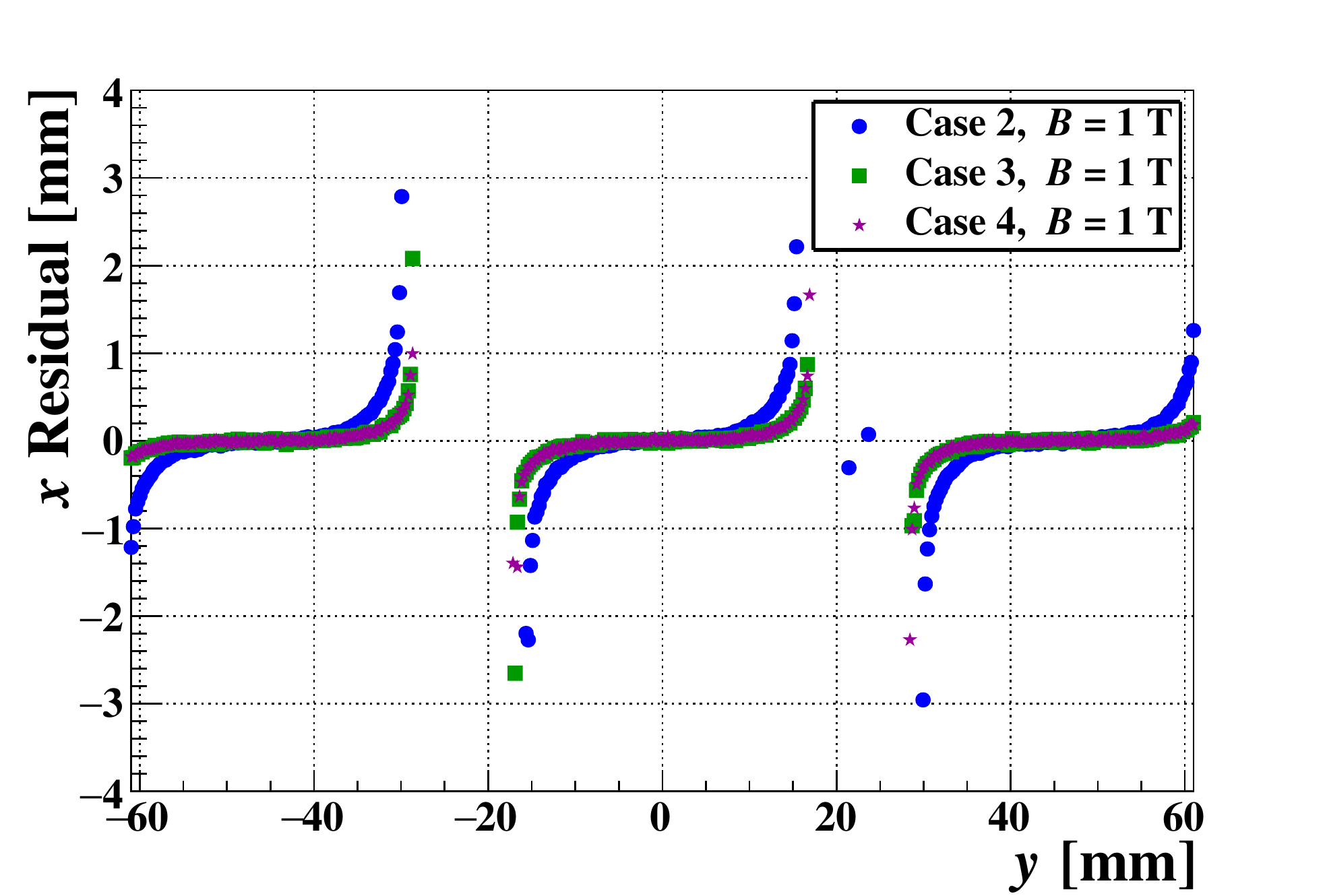}}
\subfigure[]
{\label{Study2-YHist}\includegraphics[scale=0.375]{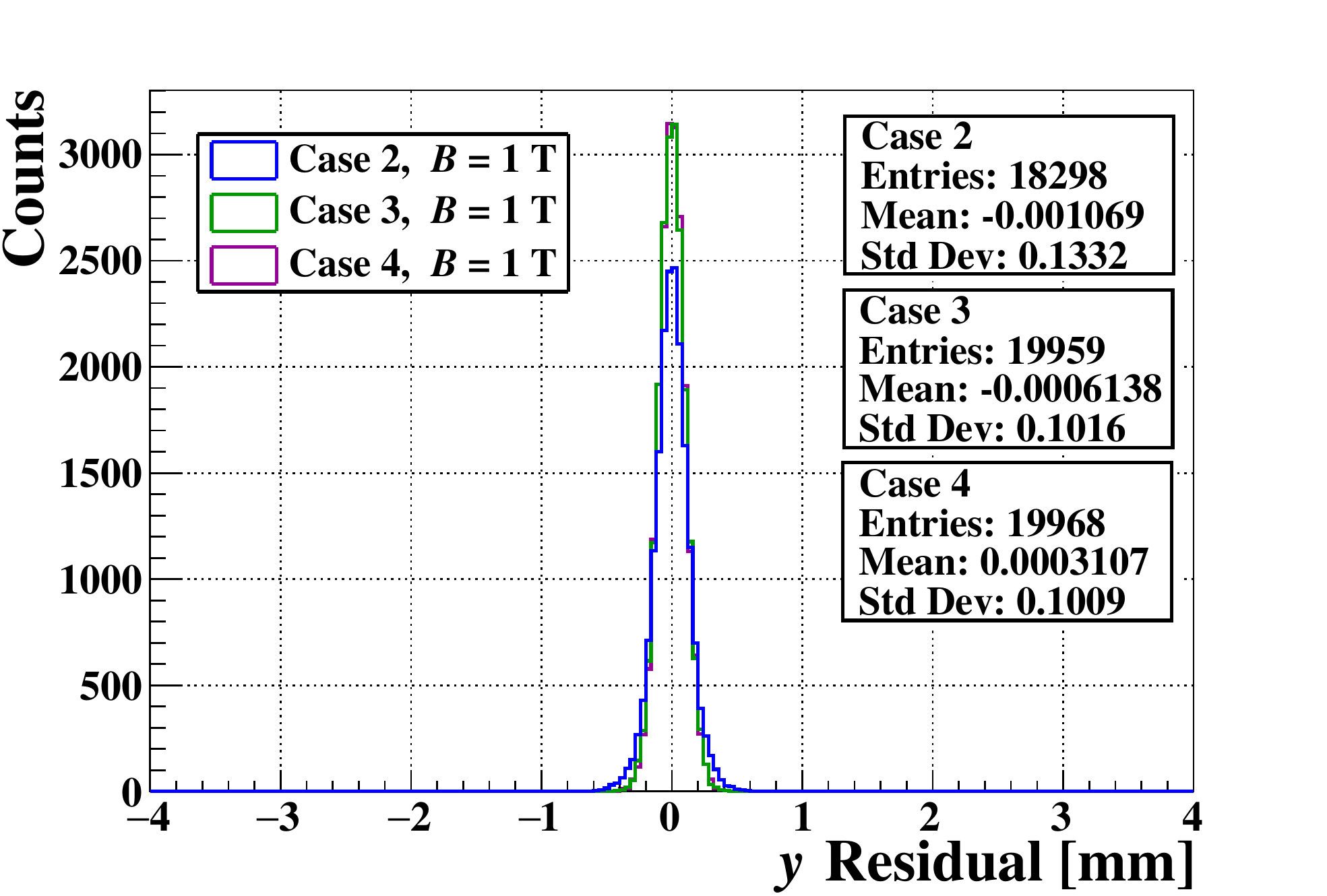}}
\subfigure[]
{\label{Study2-YRes}\includegraphics[scale=0.375]{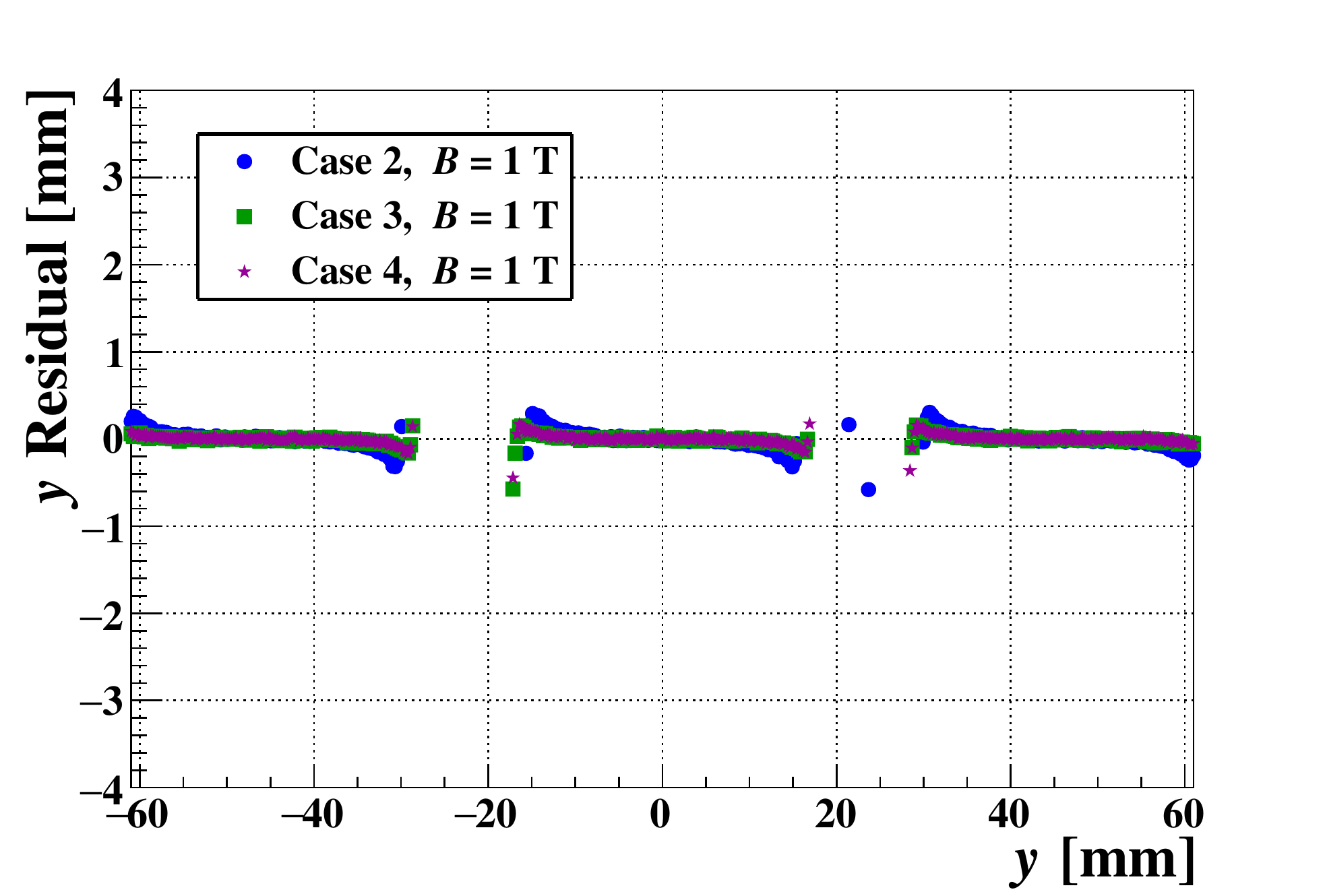}}
\caption{Comparison of (a) {\it{x}} (b) {\it{y}} residual histogram and comparison of (c) {\it{x}} (d) {\it{y}} residual for various potential applied to the copper frame. }
\label{Study2Residue}
\end{figure}

\begin{figure}[hbt]
\centering
\includegraphics[scale=0.5]{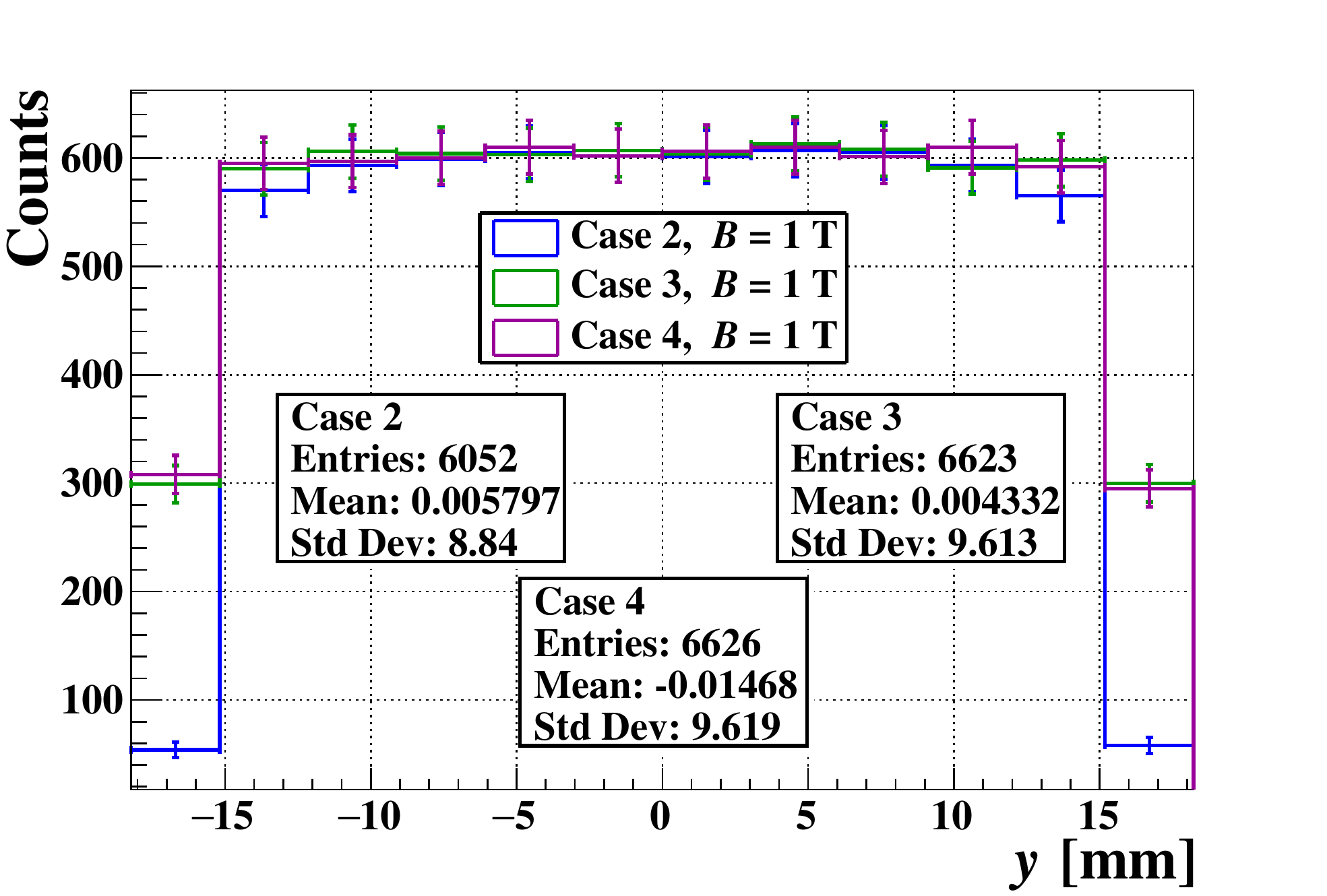}
\caption{Variation of counts along {\it{y}}.}
\label{PadRowVsCount-Modified}
\end{figure}

\begin{figure}[hbt]
\centering
\subfigure[]
{\label{SpatialX-New}\includegraphics[scale=0.375]{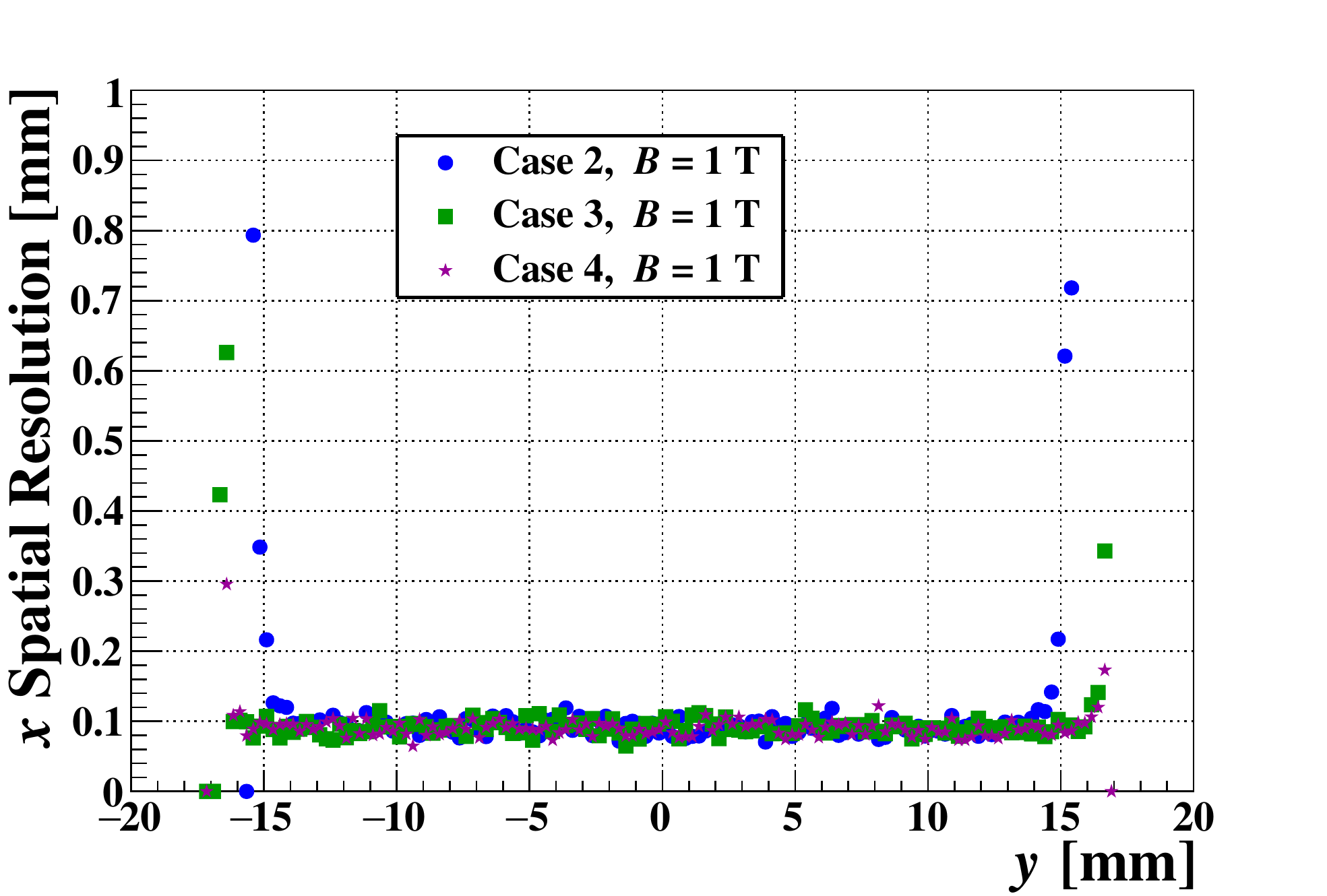}}
\subfigure[]
{\label{SpatialY-New}\includegraphics[scale=0.375]{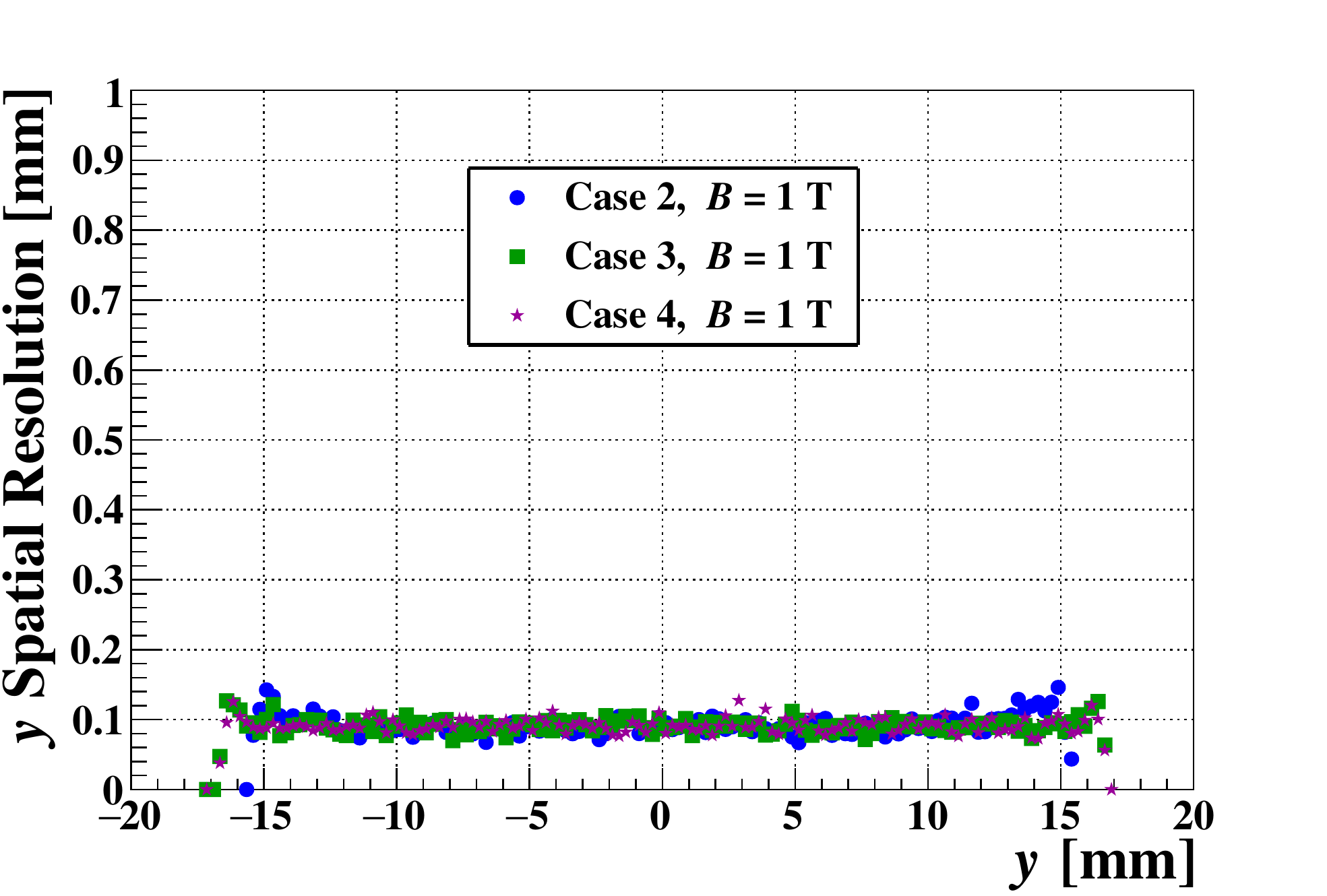}}
\caption{(a) {\it{x}} and (b) {\it{y}} spatial resolution for various potential applied to the copper frame.}
\label{Spatial-New}
\end{figure}

From Figs.~\ref{Study1-EY-Run2}, \ref{Study1-EY-Run3}, \ref{Study1-EZ-Run2} 
and \ref{Study1-EZ-Run3}, it is observed that introduction of the additional 
photoresist block used to separate anode and the copper frame does not 
introduce significant difference in the field configuration.
The effect on the computed residual, as shown in Fig.~\ref{Study1Residue} is, as a result, negligible.
On the other hand, as shown in Fig.~\ref{Study2}, both the 
transverse and vertical fields are significantly altered from the original 
configuration when Case 3 and Case 4 are considered.
The magnitudes of distortions are reduced to approximately $50\%$ of their 
earlier values.
The two cases where the mesh and copper frame and supports are at the same 
voltages, the field configurations are naturally the same.
This fact is observed also the pattern of the residuals as shown in 
Fig.~\ref{Study2Residue} in which, just as in the case of field 
configuration, the magnitude of the residuals are reduced by more than $50\%$ 
of their original values.

The variation of the count along the track has been presented in 
Fig.~\ref{PadRowVsCount-Modified}. 
In the two cases where the mesh and copper frame and supports are 
at the same voltages, more number of electrons are focused towards the 
readout plane. 
As a result, collection of electrons at the first pad increases improving 
from close to 0 to almost $50\%$ of the maximum possible count of 600. 
The spatial resolution near the edges improves as shown in Fig.~\ref{Spatial-New}.

\section{Conclusion}

Following the experimental activities and related data analysis, we have
investigated the origin of the track distortions observed close to the
module edges.
We have been able to numerically simulate the observed patterns 
successfully, and achieved quantitative agreement with the experimental data.
This is despite the fact that the entire simulation is done with a number of simplifications in the geometry of the detectors.
The intrinsic parameters, like amplification gap, thickness of the
photoresist and the ground frame, the inter-modular distances and the TPC gas
are maintained true to the experimental values for this calculation.
The miniaturization of the detector modules in ${\it{x}}-\it{{y}}$ dimensions are made only
to avoid computational delay and complexities.

Our computations indicate that the inhomogeneity of the electric field close
to the module edges leads to a loss of efficiency of few pads close to the
edge. 
This inhomogeneity also leads to the distortion in residual as observed in 
the experimental data and degradation of spatial resolution near the 
module edges.
The presence of magnetic field complicates the matter through the
$\vec{{E}} \times \vec{{B}}$ effect, since the two fields
are no more parallel (due to the fact that the ${\it{E}}$-field is non-uniform).
Interestingly enough, the effect of magnetic field is to improve the 
different figures of merit such as efficiency and resolution.
The nature and the magnitude of the distortion closely match the results as
seen in the Micromegas based LPTPC.
The obtained agreements encourage us to continue with the study
and, if possible, propose module design modifications that can alleviate the
problem.
Several such modification has been discussed in some details.
It has been shown that it is possible to reduce the effects of distortion by maintaining 
the potential of the copper frame at a value similar to that of the micro-mesh.

This work can be considered as the initial step towards design optimization of
these complex devices. 
As a result, it opens up several areas of exploration which need to be 
investigated in future studies.
For example, variation of the shape and dimension of cover electrodes and 
copper frame need to be pursued in order to optimize the device geometry. 
Moreover, there are effects of important physical processes such as charging 
up and finite resistivity that have been ignored in this work. 
Inclusion of these processes will not only be challenging computationally, 
but will also enrich our understanding of related device physics.

\acknowledgments

This work has partly been performed in the framework of the RD51 Collaboration.
We happily acknowledge the help and suggestions of the member of the RD51
Collaboration.
We thank our collaborators from LCTPC collaboration for their help and
suggestions.
We are thankful to our respective Institutions for providing us with
the necessary facilities and IFCPAR/ CEFIPRA (Project No. 4304-1) for partial
financial support.
Finally, we thank the reviewer for helping us to improve the quality of 
the manuscript significantly.

\end{document}